\documentclass[letterpaper]{article} 
\usepackage{aaai2026}  
\usepackage{times}  
\usepackage{helvet}  
\usepackage{courier}  
\usepackage[hyphens]{url}  
\usepackage{graphicx} 
\usepackage{natbib}  
\usepackage{caption} 
\usepackage{algorithm}
\usepackage{algorithmic}
\usepackage{xcolor}
\usepackage[dvipsnames]{xcolor}
\usepackage{amsmath}
\usepackage{mathtools}
\usepackage{lipsum} 
\usepackage{subcaption}
\usepackage{caption}
\usepackage{amssymb}
\usepackage{booktabs}
\usepackage{newfloat}
\usepackage{listings}
\DeclareCaptionStyle{ruled}{labelfont=normalfont,labelsep=colon,strut=off} 
\floatstyle{ruled}
\newfloat{listing}{tb}{lst}{}
\floatname{listing}{Listing}
\title{SteerMusic: Enhanced Musical Consistency for Zero-shot
Text-guided and Personalized Music Editing}
\author{
Xinlei Niu\textsuperscript{\rm 1}\thanks{Work done during the internship at Sony AI.}, Kin Wai Cheuk\textsuperscript{\rm 2}, Jing Zhang\textsuperscript{\rm 1}, Naoki Murata\textsuperscript{\rm 2}, Chieh-Hsin Lai\textsuperscript{\rm 2}, Michele Mancusi\textsuperscript{\rm 3}, Woosung Choi\textsuperscript{\rm 2}, Giorgio Fabbro\textsuperscript{\rm 3}, Wei-Hsiang Liao\textsuperscript{\rm 2}, Charles Patrick Martin\textsuperscript{\rm 1}, Yuki Mitsufuji\textsuperscript{\rm 2}\\
}

\affiliations{
    \textsuperscript{\rm 1}Australian National University, Canberra, Australia; \\
    \textsuperscript{\rm 2} Sony AI, Tokyo, Japan; \\
    \textsuperscript{\rm 3} Sony Europe B.V., Stuttgart, Germany
    
    xinlei.niu@anu.edu.au
}

\usepackage{bibentry}

\begin{document}

\maketitle

\begin{abstract}
 Music editing is an important step in music production, which has broad applications, including game development and film production. Most existing zero-shot text-guided editing methods rely on pretrained diffusion models by involving forward-backward diffusion processes. However, these methods often struggle to preserve the musical content. Additionally, text instructions alone usually fail to accurately describe the desired music. In this paper, we propose two music editing methods that improve the consistency between the original and edited music by leveraging score distillation. The first method, \textit{SteerMusic}, is a coarse-grained zero-shot editing approach using delta denoising score. The second method, \textit{SteerMusic+}, enables fine-grained personalized music editing by manipulating a concept token that represents a user-defined musical style. \textit{SteerMusic+} allows for the editing of music into user-defined musical styles that cannot be achieved by the text instructions alone. Experimental results show that our methods outperform existing approaches in preserving both music content consistency and editing fidelity. User studies further validate that our methods achieve superior music editing quality. 
\end{abstract}
\begin{links}
    \link{Code}{https://github.com/sony/steermusic}
    \link{Demonstration page}{https://steermusic.pages.dev/}
    \link{Extended version}{https://arxiv.org/abs/2504.10826}
\end{links}

\section{Introduction}~\label{sec:intro}

Text-guided diffusion probabilistic models (DPMs)~\cite{ho2020denoising,lai2025principles} have shown impressive performance in generating diverse high-quality audio samples, including music, speech, and sounds~\cite{mariani2023multi,zhang2024speaking,niu2024soundlocd}. These text-to-audio (TTA) diffusion model trained with large scale datasets that can generate diverse samples conditioned on the natural language prompts specified. Consequently, the text-guided music editing task was proposed, which edits music  by modifying the corresponding text prompts of the source music. Unlike controllable music generation, music editing task modifies an existing piece of music, which has two primary objectives: preserving original musical content and ensuring alignment between the edited music and the desired target.

\begin{figure}[t]
  \centering
  \includegraphics[width=\linewidth]{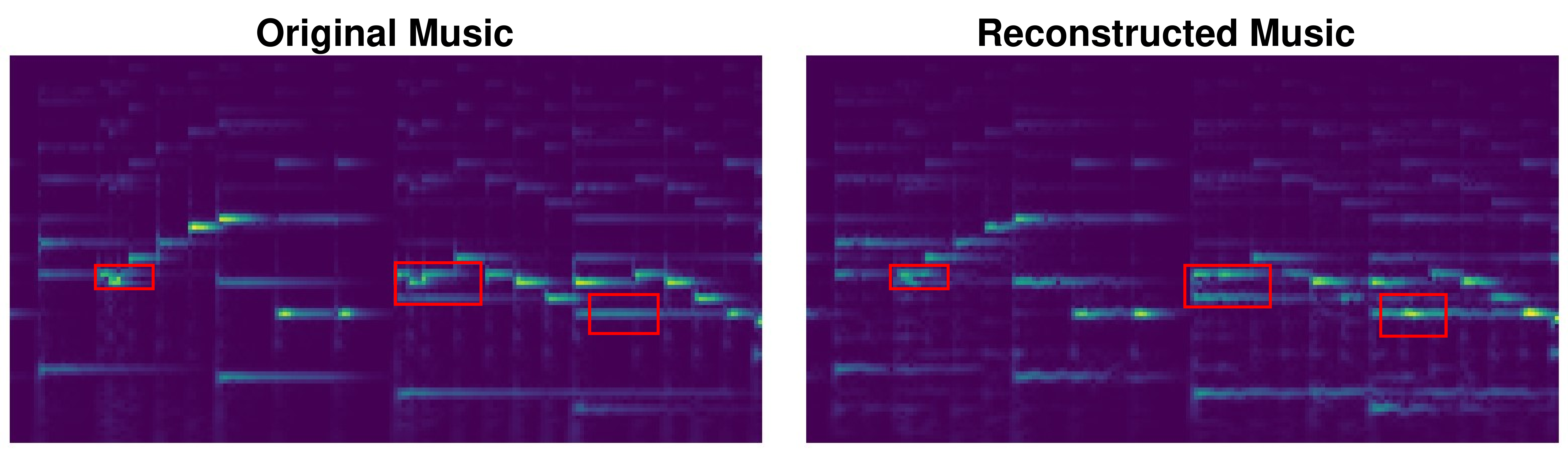}
  \caption{The distortion of the reconstructed melody (CQT1-PCC=0.721) after only 20 DDIM inversion steps.} \label{fig:drift_example}
\end{figure}

Existing music editing methods focus on training music editing models from scratch~\cite{copet2023simple} or fine-tuning pretrained TTA models~\cite{zhang2024instruct}, both of which require additional datasets or computational costs. Inspired by recent advances in image editing~\cite{brooks2023instructpix2pix,huberman2024edit,hertz2022prompt,zhang2021plug}, emerging music editing methods have instead pursued zero-shot techniques to reduce computational overhead. Existing zero-shot text-guided music editing pipelines~\cite{zhang2024musicmagus,manor2024zero,liu2024medic} introduce noise into the source music during the forward diffusion process to suppress high-frequency components (e.g., timbral information) and subsequently perform editing during the denoising phase based on target guidance in the diffusion latent space. The process of retrieving a noisy latent representation from the data is commonly referred as inversion step.
Due to the imperfect diffusion inversion process, the latent representation obtained may not fully preserve the original music content. This issue becomes more serious when the source prompt used as the condition cannot accurately capture the detailed characteristics of the music input~\cite{kawar2023imagic,paissan2023audio}. We refer to the distortion in the inverted latent representation as an ``inversion error''.
Notably, this distortion occurs even during the reconstruction of only a few of DDIM inversion steps~\cite{song2020denoising}
, which alters the melodic information in the reconstructed results compared to the original music as shown in the CQT spectrogram~\cite{brown1991calculation} in Fig.~\ref{fig:drift_example}. 
In music editing, such distortion can even be compounded, resulting in a failure to preserve instruction-irrelevant content in original music.
Although methods such as textual inversion~\cite{gal2022image} have been proposed to mitigate this issue by tuning the embeddings for near-lossless audio reconstruction~\cite{niu2024soundmorpher}, there is no way to manipulate the target text prompts within the tuned source textual embeddings in editing tasks. A more promising solution to avoid inversion error is the delta denoising score (DDS)~\cite{hertz2023delta}, a score distillation method that performs editing directly in the data space, which defines a differentiable function rendering the source input. DDS computes the difference in denoising scores between the source and target prompts through a single forward step. This approach eliminates the dependency of a full or partial forward diffusion process that could introduce inversion errors. By operating in the data space, DDS enables high-fidelity editing while preserving instruction-irrelevant content on the input.

\begin{figure}[t]
  \centering
  \includegraphics[width=\linewidth]{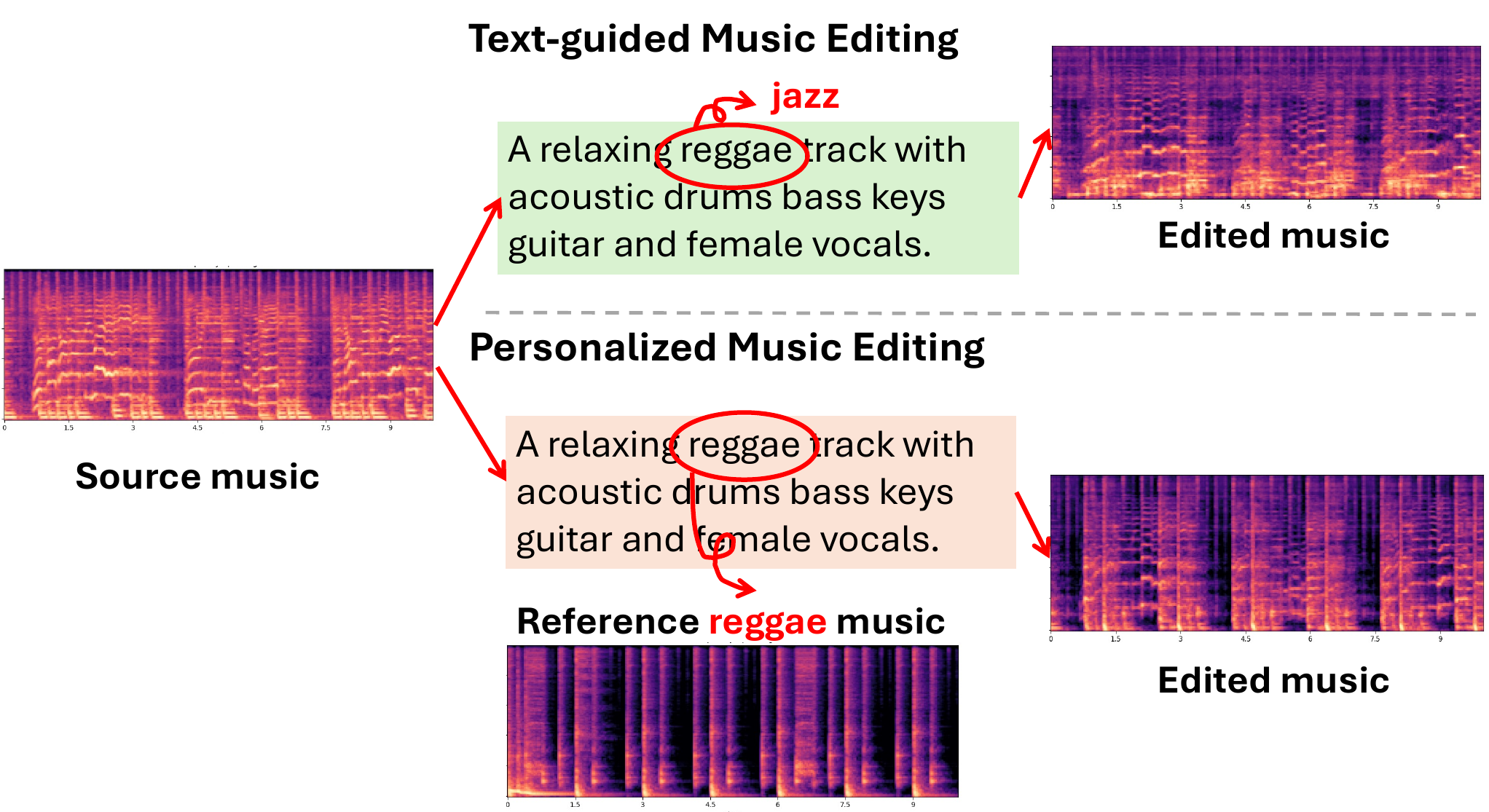}
  \caption{SteerMusic: Steering the music style with text-guided music editing or personalized music editing.}
  \label{fig:teaser}
\end{figure}

Text-guided editing enables flexible and intuitive modifications, requiring users to provide only an arbitrary text instruction to perform the desired edit. However, one limitation of text-guided music editing is that it lacks fine-grained control over the direction and nuance of editing.
For instance, editing a guitar performance into the one played by person A with a guitar brand B. Text-based editing alone struggles to specify the exact ``guitar'' required. Moreover, person A and guitar brand B might be unseen concepts for the models, rendering attempts to specify these words in the text prompt ineffective.
To enhance user-personalized controllability in text-to-music generation, DreamSound~\cite{plitsis2024investigating} introduces a pioneering approach that adapts image personalization techniques~\cite{gal2022image,ruiz2023dreambooth} to the music domain, allowing the extraction of user-defined musical characteristics from reference audio. 
Besides, DreamSound also demonstrates the potential of leveraging the personalization techniques to perform personalized music editing by manipulating learned musical concepts on the noisy source latent through the denoising process of a personalized diffusion model.
Despite their attempts, DreamSound still suffers from the inversion error, and struggles to preserve music content while editing the specific concept given in the text prompt.

In this paper, we propose two music editing methods, \textit{SteerMusic} and \textit{SteerMusic+}, that can be easily adapted to existing text-to-music DPM based on the score distillation technique. We summarize our key contributions as follows.
\begin{enumerate}
    \item We propose \textit{SteerMusic}, a zero-shot text-guided music editing pipeline based on a DDS framework, which focuses on coarse‐level editing, producing high‐fidelity results while preserving source music contents. 
    
    \item We propose \textit{SteerMusic+}, a personalized music editing method that leverages user-defined musical concepts to enable customized editing. SteerMusic+, an extension of SteerMusic, enables editing results that are not attainable through text prompts alone. For example, from reggae to the customized reggae given the reference as in Fig.~\ref{fig:teaser}.

    \item We provide extensive experiments to demonstrate that the proposed methods produce superior editing results compared to the existing state-of-the-art methods in terms of musical consistency and edit fidelity.
\end{enumerate}

\begin{figure*}[t]
  \centering
  \includegraphics[width=\linewidth]{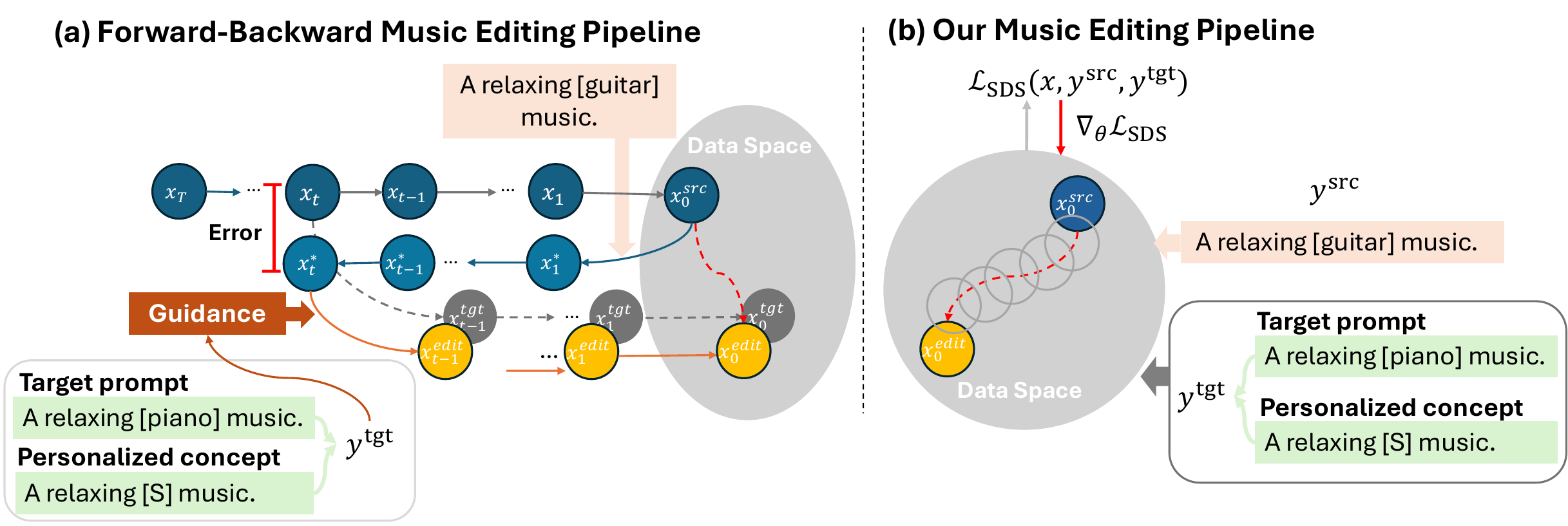}
   \caption{Overview of two music editing pipelines: (a) shows the conventional approach, which performs editing during denoising after an inversion process in the diffusion latent space; (b) shows our solution, which directly edits in data space by optimizing the differentiable function $x = g(\theta)$. The differentiable function is initialized with $x^{\text{src}}$. [$S$] denotes a user-defined concept token, and gray circles represent the optimization trajectory from source to target.}
  \label{fig:drift}
\end{figure*}

\section{Related work}\label{sec:related_work}

\subsection{Text-guided Music Generation and Editing} 
Earlier music generation work focused on low-level control signals with strict temporal alignment, such as lyrics~\cite{yu2021conditional,dhariwal2020jukebox} and MIDI~\cite{yang2017midinet} conditioning. Recently, high-level semantic prompts have gained popularity~\cite{liu2023audioldm,liu2024audioldm,kundu2024emotion,huang2023make,chowdhury2024melfusion,agostinelli2023musiclm,kreuk2022audiogen,huang2023noise2music,le2024high,li2024jen,saitosoundctm}. More recent studies further explore melody prompts from reference music~\cite{copet2024simple,novack2024ditto,novack2024ditto2,hou2024editing,chen2024jen,kumari2023multi}, enabling more precise, user-driven generation.

Music editing transforms existing audio according to target conditions while preserving the original content. Earlier approaches train models from scratch~\cite{wang2023audit,agostinelli2023musiclm,copet2023simple,hou2024editing,mariani2023multi} or fine-tune pretrained models~\cite{paissan2023audio,zhang2024instruct,tsai2024audio,han2023instructme,liu2023m}. Recent work explores zero-shot methods~\cite{zhang2024musicmagus,manor2024zero,liu2024medic} using pretrained TTA generator. However, these pipelines rely on a forward-backward diffusion process, which can introduce inversion errors as in Fig.~\ref{fig:drift} (a).

\subsection{Personalized Music Generation and Editing}
DreamSound~\cite{plitsis2024investigating} explores the possibility of capturing musical concepts from the given reference music using a personalization diffusion model~\cite{ruiz2023dreambooth,gal2022image,kumari2023multi}. This method allows users to generate new music samples by incorporating the captured personalized musical concept token into the text prompts.
In addition to personalized music generation~\cite{plitsis2024investigating,chen2024jen},
DreamSound further extends their method to personalized music editing.
However, we noticed that personalized music editing is still immature.
Existing methods~\cite{plitsis2024investigating} struggle to maintain the musical consistency while editing into the desired music concept captured in the reference audio, which use the same music editing pipeline illustrated in Fig.~\ref{fig:drift} (a).

\section{Preliminaries}

Score distillation refines generated samples using the score (i.e., the gradient of the log-density) from a pretrained diffusion model $\epsilon_\phi$ to enforce predefined constraints. It is typically implemented via probability density distillation, where gradients from the source diffusion model are used to iteratively refine a differentiable function until the desired outcome is achieved.
Score distillation sampling (SDS)~\cite{poole2022dreamfusion} pioneered score distillation by optimizing a differentiable function $x = g(\theta)$ to match a target prompt $y^{\text{tgt}}$, where the function $g(\theta)$ renders the source input $x$ with parameters $\theta$. It minimizes the loss
$\mathcal{L}_{\text{Diff}} = \mathbb{E}_{t,\epsilon}[ w(t) \| \epsilon_\phi (x_t, y^{\text{tgt}},t) - \epsilon \|_2^2]$, where $x_t$ is a noisy version of $x$ at time $t$. By omitting the UNet Jacobian, the gradient is
$
    \nabla_\theta \mathcal{L}_{\text{SDS}}(\phi,x=g(\theta),y^{\text{tgt}}) 
     =\mathbb{E}_{t,\epsilon}[w(t)(\epsilon_\phi(x_t,y^{\text{tgt}},t) -\epsilon)\frac{\partial x}{\partial \theta}] \label{eqn:sds}
$,
where $y^{\text{tgt}}$ is the target prompt, $x_t$
is a noisy latent of $x = g(\theta)$ at time step $t$, and $w(t)$ is a weighting function. However, SDS often produces blurry results in image editing
~\cite{wang2023prolificdreamer,hertz2023delta}. The delta denoising score (DDS)~\cite{hertz2023delta} addresses this issue by computing the delta score between the source prompt $y^{\text{src}}$ and the target prompt $y^{\text{tgt}}$. 

In image editing, DDS refines only regions relevant to the target prompt $y^{\text{tgt}}$, preserving the rest of the image. Given source input $x^{\text{src}}$ with prompt $y^{\text{src}}$ and target prompt $y^{\text{tgt}}$, the gradient over $\theta$ is
\begin{multline}
    \nabla_\theta \mathcal{L}_{\text{DDS}}(\phi,x=g(\theta),y^{\text{tgt}},x^{\text{src}},y^{\text{src}}) \\
    = \mathbb{E}_{t,\epsilon}[w(t)(\epsilon_\phi(x_t,y^{\text{tgt}},t)  -\epsilon_\phi(x^{\text{src}}_t,y^{\text{src}},t))\frac{\partial x}{\partial \theta}] \label{eqn:dds}
\end{multline}
where $x_t$ and $x_t^{\text{src}}$ share the same sampled noise $\epsilon$ and the timestep $t$. 

Although different variants of SDS and DDS have been proposed for the image domain~\cite{yu2025dreamsteerer,hertz2023delta,nam2024contrastive,lin2025zero}, the application of DDS in music editing remains underexplored. In the next section, we will explain how to incorporate DDS into music editing tasks.

\section{Method}
In this section, we introduce two music editing methods: \textit{SteerMusic} for zero-shot text-guided editing, and \textit{SteerMusic+} for personalized editing using user-defined concepts from reference music. Unlike the forward-backward pipeline in Fig.\ref{fig:drift} (a), our methods edit directly in a data space (Fig.\ref{fig:drift} (b)), yielding better musical content consistency.

\subsection{SteerMusic: Zero-shot Text-guided Music Editing}~\label{sec:SteerMusic}
\begin{figure*}[t]
  \centering
  \includegraphics[width=\linewidth]{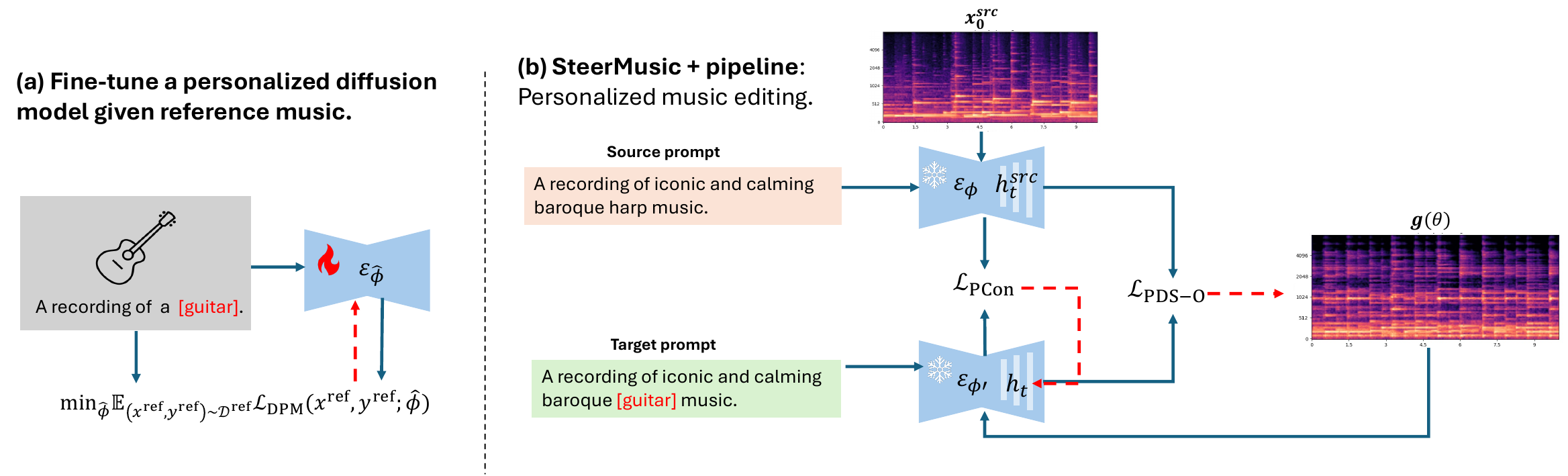}
  \caption{Overview of the SteerMusic+ pipeline: (a) Personalized diffusion model (PDM) fine-tuned using $\mathcal{D}^{\text{ref}}$ and a user-defined [guitar] concept token. (b) Personalized editing using the PDM $\epsilon_{\phi'}$ from (a).  Red dashed lines indicate gradient flows.}
  \label{fig:SteerMusic+}
\end{figure*}

We introduce \textit{SteerMusic}, a zero-shot text-guided music editing method that performs editing in the data space. In this setting, our goal is to edit a source music signal $x^{\text{src}}$ by modifying its corresponding text prompt $y^{\text{src}}$, which $y^{\text{src}}$ is a brief description of the source music that includes the specific musical attribute intended for modification. The modified text prompt $y^{\text{tgt}}$ acts as the target prompt to guide the editing.
To obtain desirable results, the musical content shared by $y^{\text{tgt}}$ and $y^{\text{src}}$ should be preserved, changing only the content that is distinct in $y^{\text{tgt}}$. For example, if the only change in $y^{\text{tgt}}$ compared to $y^{\text{src}}$ is replacing only the word ``piano'' with ``guitar'' while keeping the rest of the sentence unchanged, the edited music should preserve the melody and tempo, while modifying only the musical instrument.

To achieve this goal, we adopt the DDS, which has been previously explored only in the image domain. To the best of our knowledge, we are the first to investigate the potential application in music editing. Following DDS, we define $x = g(\theta)$ rendering a source music signal $x^{\text{src}}$. We picked $g(\theta)=\theta$ as the differentiable function in Eq.~\ref{eqn:dds}, where we initialize $\theta =x_0^{\text{src}}$. i.e. $x=x_0^{\text{src}}$. 
In SteerMusic, we set $\epsilon_\phi$ as a pretrained TTA or text-to-music DPMs.

Similar to~\citet{hertz2023delta}, the delta score in Eq.~\ref{eqn:dds} steers the optimization process toward the target prompt while reducing the noisy editing direction commonly associated with vanilla SDS, leading to enhanced edit fidelity.
SteerMusic method enables flexible editing using text instructions alone; however, it is limited to coarse-grained editing, as text prompts often lack the precision information to capture fine-grained musical details.

\subsection{SteerMusic+: Personalized Music Editing}~\label{sec:SteerMusic+}
As mentioned before, text-guided editing lacks customization for precise music editing, such as transferring music to a specific style. To enable fine-grained personalized editing, we propose 
\textit{SteerMusic+}, an extension of SteerMusic. In this setting, we have a set of source music and prompt pair $\{x^{\text{src}},y^{\text{src}}\}$. The target prompt $y^{\text{tgt}}$ is constructed by modifying $y^{\text{src}}$ to manipulate a user-defined concept token $[S]$, representing the desired customization (see example in Fig.~\ref{fig:SteerMusic+} (b)). SteerMusic+ uses two pretrained diffusion models:
\begin{itemize}
    \item A pretrained diffusion probabilistic model (DPM), denoted as $\epsilon_{\phi}$, which serves as a reference for maintaining consistency with the original music content; and
    \item A personalized diffusion model (PDM), denoted as $\epsilon_{\phi '}$, is fine-tuned on a small set of data containing reference music to capture the user-defined concept $[S]$ and guide the editing process toward the desired direction.
\end{itemize}
We define successful personalized editing by the criteria:
\begin{itemize} 
\item The instruction-irrelevant part (i.e., \{$y^{\text{tgt}}\cap y^{\text{src}}$\}) in the source music should be maintained.
\item The edited musical attributes should perceptually align with the intended personalized musical concept $[S]$.
\end{itemize}
We now present the SteerMusic+ method that enables personalized music editing with enhanced music consistency.

\textbf{Personalized Diffusion Model (PDM)} serves a foundational part in SteerMusic+.
Since training a PDM has been extensively studied in~\citet{plitsis2024investigating,ruiz2023dreambooth,gal2022image,kumari2023multi}, we assume the availability of a pretrained text-to-music PDM in SteerMusic+, as SteerMusic+ is a plug-in pipeline compatible with existing PDMs.
The text-to-music PDM $\epsilon_{\phi '}$ captures the user-defined musical concept $S$ by fine-tuning a pretrained DPM $\epsilon_{\phi}$ under a small set of reference music $\mathcal{D}^{\text{ref}} = \{(x^{\text{ref}},y^{\text{ref}})_n\}^N_{n=1}$, which $N$ can be as few as $1$~\cite{plitsis2024investigating}. The fine-tuning is achieved via optimizing the objective
\begin{equation}
    \phi ' \in \text{arg min}_{\hat{\phi}}  \mathbb{E}_{(x^{\text{ref}},y^{\text{ref}})\sim\mathcal{D}^{\text{ref}}} \mathcal{L}_{\text{DPM}}(x^{\text{ref}},y^{\text{ref}};\hat \phi) \label{eqn:pdm}
\end{equation}
where $\hat{\phi}$ is initialized with a pretrained DPM weights $\phi$. As illustrated in Fig.~\ref{fig:SteerMusic+} (a), the prompt $y^{\text{ref}}$ takes form of ``a recording of a [$S$]'', where the placeholder [$S$] corresponds to a defined new concept word embedding. During inference, the PDM can generate music with the newly learned concept (e.g., ``A disco song with a [$S$]''). 
The dataset $\mathcal{D}^{\text{ref}}$ consists of reference audio clips that encapsulate the concept users aim to extract. According to~\citet{plitsis2024investigating}, the concept represents a musical style, which can be either an instance of instrument sounds or a specific genre that cannot be yielded even with the most detailed textual description. For instance, if the objective is to capture the user's guitar playing style, the reference audio should feature performances on the user's guitar playing. Conversely, if the goal is to capture the concept of jazz, the reference audio should consist of recordings that exemplify the same jazz style.

While the PDM can generate new music based on concepts from $\mathcal{D}^{\text{ref}}$, it cannot directly edit existing music using $y^\text{tgt}$ that contains the concepts. Next, we introduce key components of SteerMusic+ that enable personalized editing with PDMs.

\textbf{Personalized Delta Score (PDS)} is an extension of Eq.~\ref{eqn:dds} to enable personalized music editing. We define the PDS loss as $\mathcal{L}_{\text{PDS}}(\phi ', \phi, x=g(\theta) ,y^{\text{tgt}},x^{\text{src}},y^{\text{src}}) = \mathbb{E}_{t,\epsilon} [w(t)\|\epsilon_{\phi '}(x_t,y^{\text{tgt}},t) - \epsilon_{\phi}(x^{\text{src}}_t,y^{\text{src}},t) \|^2_2]$.
By omitting the UNet Jacobian, the gradient over $\theta$ is given by
$
   \nabla_\theta \mathcal{L}_{\text{PDS}} = \mathbb{E}_{t,\epsilon}[w(t)(\epsilon_{\phi '}(x_t,y^{\text{tgt}},t) -\epsilon_{\phi}(x^{\text{src}}_t,y^{\text{src}},t))\frac{\partial x}{\partial \theta}] 
$,
where $x_t$ and $x_t^{\text{src}}$ share the same sampled noise $\epsilon$ at the time step $t$.
This modified delta score can be decomposed into two components: the score of $\epsilon_{\phi '}$ provides the desired direction to guide the editing to match the target prompt with the concept $[S]$. The score of $\epsilon_{\phi}$ reduces the noisy direction of unintended modification areas.
The delta score between the PDM $\epsilon_{\phi '}$ and the DPM $\epsilon_{\phi}$ may not produce an effective direction toward $y^\text{tgt}$, as $\epsilon_{\phi '}$ shifted to the reference distribution $\mathcal{D}^{\text{ref}}$. We introduce an additional component to compensate for the distribution shift induced by the score of $\epsilon_{\phi '}$.

\textbf{Distribution Shift Regularization.}
To bridge the distribution gap between $\epsilon_{\phi '}$ and $\epsilon_{\phi}$, we introduce a regularization term to regularize the edited score to the personalized diffusion model $\epsilon_{\phi '}$. We wish to minimize the distribution shift between two diffusion models by adding a constraint as
\begin{equation}
\begin{aligned}
\min_{\theta}\;& \mathcal{L}_{\text{PDS}}(\phi ',\phi,x=g(\theta),y^{\text{tgt}},x^{\text{src}},y^{\text{src}}),\\
\text{subject to}\;& \mathcal{L}_{\text{shift}}(\phi ',\phi,x=g(\theta),y^{\text{tgt}}) - \zeta \leq 0.
\end{aligned}
\end{equation}
where $\zeta$ is a small amount of constant; the regularization term is 
$\mathcal{L}_{\text{shift}}(\phi ',\phi,x=g(\theta), y^{\text{tgt}}) = \mathbb{E}_{t,\epsilon}[w(t) \|\epsilon_{\phi '}(x_t, y^{\text{tgt}},t) - \epsilon_{\phi}(x_t,y^{\text{tgt}},t) \|^2_2 ]$.
The gradient in respect to $\theta$ is given by
$
\nabla_\theta \mathcal{L}_{\text{shift}} =  \mathbb{E}_{t,\epsilon}[w(t)(\epsilon_{\phi '}(x_t,y^{\text{tgt}},t) - \epsilon_{\phi}(x_t,y^{\text{tgt}},t))\frac{\partial x}{\partial \theta} ]
$.
Therefore, the overall gradient through $\theta$ is 
\begin{multline}~\label{eqn:psd_0}
    \nabla_\theta \mathcal{L}_{\text{PDS-O}}(\phi ',\phi,x=g(\theta),y,x^{\text{src}},y^{\text{src}}) 
    \\ =  \underbrace{\nabla_\theta \mathcal{L}_{\text{PDS}}(\phi ',\phi,x=g(\theta),y^{\text{tgt}},x^{\text{src}},y^{\text{src}})}_{\text{`Delta score points to edit direction'}} \\
    + \lambda  \underbrace{\nabla_\theta  \mathcal{L}_{\text{shift}}(\phi ',\phi,x=g(\theta),y^{\text{tgt}})}_{\text{`Delta score regularizes distribution shift'}}
\end{multline}
where $\lambda$ is a constant that adjusts regularization strength. 

\textbf{Personalized Contrastive (PCon) Loss.}
Eq.~\ref{eqn:psd_0} formulates a regularized reference guided editing direction, where the regularized delta score encourages alignment with the target prompt while mitigating the distribution shift.
However, it does not explicitly enforce the fidelity to the concept, which may lead to suboptimal editing quality.
To further enhance the fidelity of the edit, 
we incorporate a PCon loss between temporal features, which is modified from a patch-wise contrastive loss~\cite{nam2024contrastive}. PCon loss extracts intermediate features $h^{\text{src}}_l$ and $h_l$ that pass through the residual block and the self-attention block from $\epsilon_{\phi}$ conditioned on $y^{\text{src}}$ and $\epsilon_{\phi '}$ conditioned on $y^{\text{tgt}}$ at the $l$-th self-attention layer, respectively. The features are then reshaped to size $\mathbb{R}^{T_l \times F_l \times C_l}$, where $T_l,F_l$, and $C_l$ represent the size of the temporal, spatial, and channel dimensions in the $l$-th layer, respectively. The patch corresponding to the temporal location on the feature map $h^{\text{src}}_l$ is designated as `positive', and vice versa. The PCon loss is defined as
\begin{equation}
    \mathcal{L}_{\text{PCon}}(x,x^{\text{src}}) = \mathbb{E}_{h} [\sum_l \sum_{t'} \ell(h_l^{t'},h_l^{\text{src},t'},h_l^{\text{src},T_l\backslash t'})] \label{eqn:patch_personal}
\end{equation}
\begin{equation}
    \ell(h,h^+,h^-) = -\text{log}(\frac{\text{exp}(h\cdot h^+/\tau)}{\text{exp}(h\cdot h^+/\tau)+ \text{exp}(h\cdot h^-/\tau)})  \nonumber
\end{equation}
where $t' \in \{1,...,T_l\}$ represents the temporal location query patch, the positive patch as $h_l^{src,t'}$ while the other patches as $h_l^{src,T_l\backslash t'}$. $\text{exp}(h\cdot h^+/\tau)$ is a a positive sample with the same temporal location, $\text{exp}(h\cdot h^-/\tau)$ is a negative sample with a mismatched temporal location in the self-attention features, $\tau$ is a temperature parameter 
as $\tau >0$.

The gradient of $\mathcal{L}_{\text{PCon}}(x,x^{\text{src}})$ will propagate to the hidden state of self-attention layers $h$ in personalized diffusion $\epsilon_{\phi '}$. Given that the personalized diffusion $\epsilon_{\phi '}$ has a distribution shift over the reference dataset $\mathcal{D}^{\text{ref}}$, the $\mathcal{L}_{\text{PCon}}$ explicitly encourages feature similarity at the frequency domain in self-attentions, particularly for attributes that distinguish the target concept. This reinforcement leads the model to prioritize concept consistency over strict temporal alignment with the source music. Consequently, $\mathcal{L}_{\text{PCon}}$ amplifies the distinctive characteristics of the target concept of $\epsilon_{\phi '}$ in SteerMusic+, ensuring that the edited music maintains stronger fidelity to the desired style while allowing structural variations.

\section{Experiments}\label{sec:exp}
\begin{table*}[t]
\centering
  {\fontsize{8pt}{9pt}\selectfont
  \begin{tabular}{c|ccccc|cc}
    \toprule
    Method  &FAD$_{\text{CLAP}}$$\downarrow$  &FAD$_{\text{Vggish}}$$\downarrow$ & CQT1-PCC$\uparrow$& LPAPS$\downarrow$ &  CLAP$\uparrow$&  MOS-P$\uparrow$ & MOS-T$\uparrow$\\
    \midrule    
    DDIM& \underline{0.477} & 4.713 &  0.330  & 5.377 &\textbf{0.264} & 1.37 & \underline{1.91}\\
    SDEdit & 0.638 & 6.749 & 0.169 & 6.208 & 0.218 & 0.92 & 1.68\\
    MusicMagus & 0.593  & 7.631 & \underline{0.338}& \underline{5.243} & 0.238  & \underline{2.11} &  1.57\\ 
    ZETA  & 0.509  &   \underline{3.380} & 0.293 & 5.458 & 0.252  & 1.22 & 1.60\\
    SteerMusic& \textbf{0.278} & \textbf{2.426} & \textbf{0.480}  & \textbf{3.772} & \underline{0.259} & \textbf{2.92} & \textbf{2.50} \\
    \bottomrule
  \end{tabular}
  }
  \caption{Model comparison on zero-shot text-guided music editing task using the ZoME-Bench dataset.}\label{tab:SteerMusic}
\end{table*}
\subsection{Evaluation Metrics}

We evaluate music editing objectively based on two aspects: musical consistency (content preservation before and after editing) and editing fidelity. We follow~\citet{manor2024zero} and calculate the following objective metrics for measurement. 
To evaluate \textit{musical consistency}, we use:
\begin{itemize}
    \item \textbf{Fréchet Audio Distance (FAD)}~\cite{kilgour2018fr} which measures the distributional difference between source and edited music (lower the better). We calculate FAD based on both VGGish~\cite{hershey2017cnn} and clap-laion-music~\cite{wu2023large} embeddings, denoted as \textit{FAD$_{\text{Vggish}}$} and \textit{FAD$_\text{CLAP}$} respectively.
    \item \textbf{LPAPS}~\cite{iashin2021taming}, an audio version of LPIPS~\cite{zhang2018unreasonable}, which quantifies the consistency of the edited audio relative to the source audio (lower the better).
    \item \textbf{Top-1 Constant-Q Transform Pearson Correlation Coefficient (CQT1-PCC)}, which measures melody consistency between the source and edited music (higher the better). The CQT1-PCC~\cite{brown1991calculation} extracts the main melody of the music audio and has been shown to outperform traditional chroma-based features in representing melodic characteristics~\cite{hou2024editing}. While existing metrics such as FAD and LPAPS provide insight into audio quality and perceptual similarity, they fall short in comprehensively capturing melodic structure. Moreover, existing transcription models~\cite{cwitkowitz2024timbre,bittner2022lightweight,gardner2022mt,chang2024yourmt3+,mancusi2025latent} effectively extract melodies from real music, they are unreliable for synthesized audio. To address these limitations, we introduce CQT1-PCC as a supplementary objective metric, specifically designed to quantify melodic consistency in generative music editing. This metric enables a more targeted evaluation of whether the core melodic structure of the source audio is retained in the edited result. 
\end{itemize}

To evaluate \textit{editing fidelity}, we use:

\begin{itemize}
    \item \textbf{CLAP Score}~\cite{wu2023large} which
measures the alignment between edited music and the target prompt in text-guided music editing (higher is better).
    \item \textbf{CDPAM}~\cite{manocha2021cdpam}, a perceptual audio metric that leverages deep learning representations to measure perceptual distance between audios such as music and speech~\cite{jacobellis2024machine,hai2024dpm,gui2024adapting}. We use CDPAM to evaluate audio perceptual similarity between reference music and the edited result in personalized music editing (lower the better).
\end{itemize}

\textbf{Subjective evaluation.} 
We designed a mean opinion score (MOS) study to evaluate target editing fidelity (MOS-T) and source content preservation (MOS-P) by asking participants to rate the results from 1-Bad to 5-Excellent~\cite{international1996methods} for randomly selected edited samples. We provide more experimental details in our extended version.

\subsection{Zero-shot Text-guided Music Editing} ~\label{sec:SteerMusic_exp}
In this part, we evaluate our SteerMusic method on the zero-shot text-guided music editing task.

\textbf{Dataset.} 
We use the ZoME-Bench dataset~\cite{liu2024medic} which includes 1,000 10-second audio samples from MusicCaps~\cite{agostinelli2023musiclm}, each paired with source and target text prompts. We evaluate our models on four well-defined editing tasks that require modifying a specific aspect of the audio while preserving the original melody: change instrument (131 clips), change genre (134), change mood (100), and change background (95).
To assess long-form editing, we use the MusicDelta subset of MedleyDB~\cite{musicdelta} with ranging from 20 seconds to 5 minutes, comprising 34 excerpts of varying styles and lengths, with prompts from~\citet{manor2024zero}.

\textbf{Baseline.}
We compare SteerMusic with zero-shot text-guided music editing methods plug-in the same pretrained AudioLDM2~\cite{liu2024audioldm}, including SDEdit~\cite{meng2021sdedit}, DDIM~\cite{song2020denoising}, ZETA~\cite{manor2024zero}, and MusicMagus~\cite{zhang2024musicmagus}. To ensure statistical reliability, experiments are conducted using multiple random seeds.
We are unable to include MelodyFlow~\cite{le2024high} and MEDIC~\cite{liu2024medic} due to the lack of source code.
We exclude AudioEditor~\cite{jia2025audioeditor} and AudioMorphX~\cite{liang2024audiomorphix}, which are designed for general sound editing rather than music.

\begin{figure}[t]
  \centering
  \includegraphics[width=\linewidth]{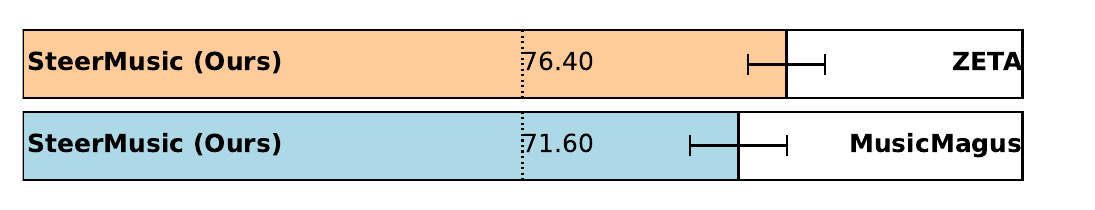}
  \caption{
  User preference for SteerMusic: percentage of users preferring our method over ZETA and MusicMagus.
  }
  \label{fig:user_study1}
\end{figure}

\begin{figure*}[t]
  \centering
  \includegraphics[width=\textwidth]{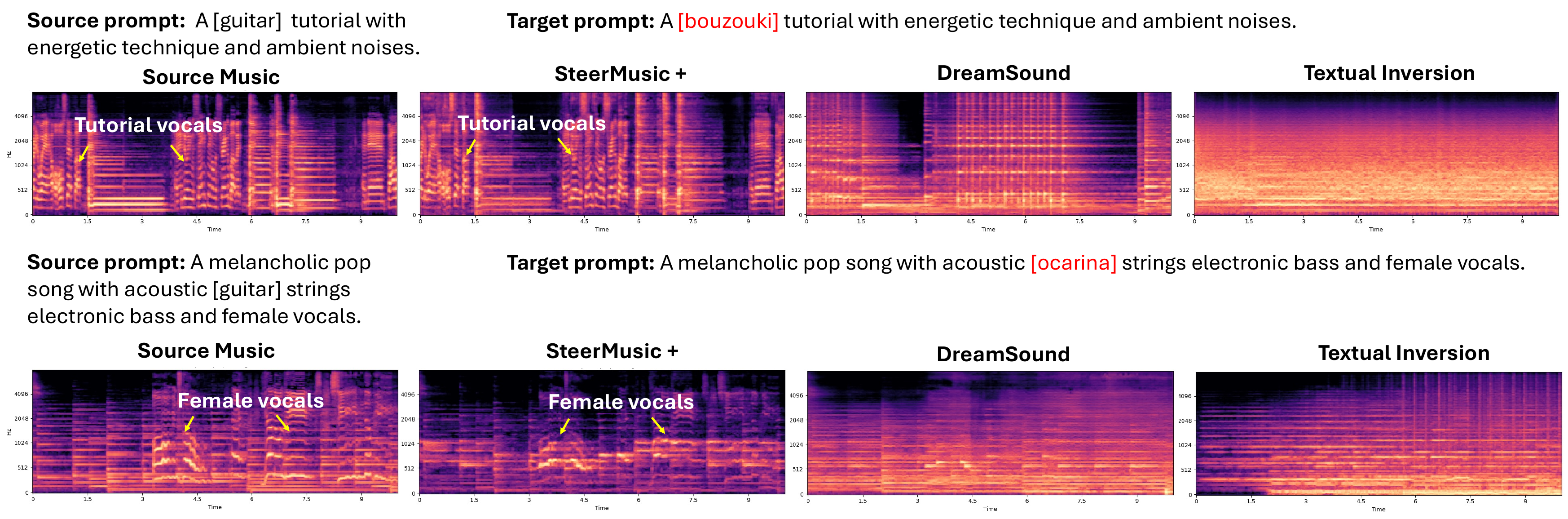}
  \caption{A visualization of edited results between SteerMusic+ and baselines in personalized music editing. \textnormal{SteerMusic+ preserves instruction-irrelevant musical content on the source music}.}
  \label{fig:compare1}
\end{figure*}

\begin{table*}[t]
  \centering
  {\fontsize{8pt}{9pt}\selectfont
  \begin{tabular}{c|ccccc|cc}
    \toprule
    Method   & FAD$_{\text{CLAP}}$$\downarrow$& FAD$_{\text{Vggish}}$$\downarrow$ & CQT1-PCC$\uparrow$ & LPAPS$\downarrow$& CDPAM$\downarrow$ &  MOS-P$\uparrow$ &  MOS-T$\uparrow$ \\
    \midrule
    Textual Inv. & 0.789 & 9.688 & 0.216 & 5.083 & 0.713 & 1.64 & 1.63\\
    DreamSound& 0.902 &10.686  & 0.292  & 5.082 & 0.609 & 1.42 & 1.81\\
    SteerMusic+ & \textbf{0.362} & \textbf{4.434}&\textbf{0.399} & \textbf{4.125} & \textbf{0.593} &\textbf{3.07} &\textbf{2.47}\\
    \bottomrule
  \end{tabular}
  }
  \caption{Model comparison on personalized music editing task using the ZoME-Bench dataset.}\label{tab:SteerMusic+_overall}
\end{table*}

\textbf{Experimental results.}
Tab.\ref{tab:SteerMusic} compares SteerMusic with zero-shot baselines across various style transfer tasks. SteerMusic achieves higher source consistency, shown by improved CQT1-PCC, lower LPAPS, and FAD. While DDIM attains a slightly higher CLAP score (+5e-3), its low CQT1-PCC and high LPAPS indicate poor preservation of source content due to lack of further source consistency constraints during denoising.
In contrast, SteerMusic effectively balances source consistency and edit fidelity, fulfilling the core objective of music editing.
Furthermore, our method attains the highest MOS scores and yields statistically significant improvements over all baseline models. An ANOVA test shows MOS-P $=2.92$ with p-value $=7.37\times10^{-27}$ and MOS-T $=2.5$ with p-value $=3.24\times10^{-7}$. These results demonstrate that SteerMusic provides substantially better editing performance in terms of both source-music consistency and editing fidelity as perceived by human listeners.

To assess real-world applicability, we further evaluate our method on MusicDelta dataset in Tab.~\ref{tab:musicdelta}, which consists of varying lengths of music clips. MusicMagus fails in this experiment as it was designed for 5-second editing and doesn't support long-form music, and hence is excluded. SteerMusic consistently outperforms other baselines in terms of edit fidelity and source consistency, demonstrating its robustness and effectiveness in handling longer and more complex music editing.

\textbf{User preference study.} We evaluate SteerMusic with the user preference study following the design in~\citet{manor2024zero}. To reduce cognitive load and improve reliability, we compare SteerMusic with the top-performing baselines, MusicMagus and ZETA.
In this study, users were asked to answer a sequence of 20 questions, each question contains original music, an editing instruction, and two edited results. Users were instructed to select the edited result that better matches the instruction while preserving the main content of the original music. We collected 25 full responses, which the participants having a minimum of 1 and average of 5 years of music training. As shown in Fig.~\ref{fig:user_study1}, our method was clearly preferred over all competing methods.

\subsection{Personalized Music Editing } ~\label{sec:SteerMusic+_exp}
In this part, we evaluate our SteerMusic+ method, which is designed for personalized music editing task.

\textbf{Dataset.}
To cover both common and exotic concepts, we selected eight representative musical concepts from the 32 defined in~\citet{plitsis2024investigating}: four instruments style (Guitar, Bouzouki, Ocarina, Sitar) and four genres (Morricone, Reggae, Hiphop, Sarabande). Each concept includes a placeholder instruction and five 10-second reference clips from YouTube and FreeSound. We use the ``change instrument'' and ``change genre'' tasks from ZoME-Bench~\cite{liu2024medic}, replacing the original target prompt with the selected concept token. Due to the lack of detailed instructions in MusicDelta, we use the standardized prompt as ``A recording of a [style] song.'', where [style] is either the source style or the target style concept [S].

\begin{table*}[t]
  \centering
{\fontsize{8pt}{9pt}\selectfont
  \begin{tabular}{l|llllllllll}
    \toprule
    Method & FAD$_{\text{CLAP}}$$\downarrow$  & FAD$_{\text{Vggish}}$$\downarrow$ & CQT1-PCC$\uparrow$ & LPAPS$\downarrow$ & CDPAM$\downarrow$ & CLAP$\uparrow$\\
    \midrule
    DDIM & 0.646 & 3.336 & 0.245 & 4.972 & - & 0.316 \\
    ZETA & 0.665 & 3.789 & 0.296 & 5.071 & - & 0.319 \\
    SDEdit & 0.818 & 8.757 & 0.137 & 5.996& - & 0.310 \\
    SteerMusic& \textbf{0.622}& \textbf{2.559} & \textbf{0.351} & \textbf{4.122} & - & \textbf{0.321} \\
    \midrule
    DreamSound & 0.847 & 8.972 & 0.220 & 5.318 & 0.583 & -\\
    SteerMusic+  & \textbf{0.666} & \textbf{5.506} & \textbf{0.273} & \textbf{4.574} & \textbf{0.581} & -\\
    \bottomrule
  \end{tabular}
  }
  \caption{Model comparison of SteerMusic and SteerMusic+ on MusicDelta dataset.}\label{tab:musicdelta}
\end{table*}

\textbf{Baseline.} We set two existing personalized music editing methods proposed by~\citet{plitsis2024investigating} as the baselines, Textual inversion and DreamSound. Textual inversion optimizes a concept embedding, whereas DreamSound fine-tunes an AudioLDM2 with rare-token identifiers~\cite{ruiz2023dreambooth}.
Both methods perform personalized music editing by manipulating the concept token during denoising process.
We follow the official codes provided by~\citet{plitsis2024investigating} to obtain text-to-music PDM. We reproduce the personalized music editing methods by calculating noisy latent representations $x_t$ from $x_0^{\text{src}}$ of a DPM conditioned on the source prompt with a predefined shallow time step $t$ using DDIM inversion~\cite{song2020denoising}, where $t=30$. We denoise $x_t$ on the PDM~\footnote{\label{foot:textinv} Textual inversion optimizes only the concept token embedding rather than fine-tuning a PDM, we denoise $x_t$ using the DPM employed during DDIM inversion. Consequently, textual inversion is incompatible with SteerMusic+ pipeline, which relies on a PDM.} conditioned on the target prompt linked to the learned concept to obtain $x_0^{\text{tgt}}$. To ensure statistical reliability, experiments are conducted using multiple random seeds. We do not compare with Jen-1 DreamStyler~\cite{chen2024jen}, which is a personalized music generation method.

\textbf{Experimental results.}
We plug SteerMusic+ into the PDM used in DreamSound\textsuperscript{\ref{foot:textinv}}.
Tab.~\ref{tab:SteerMusic+_overall}
shows that SteerMusic+ achieves superior musical consistency compared to the baselines. It indicates that the edited outputs of SteerMusic+ successfully preserve instruction-irrelevant music content in the source music. Furthermore, SteerMusic+ performs accurate edits that align well with the concepts captured from references, as indicated by the low CDPAM compared to the baseline methods. These objective evaluation results aligned with the subjective metrics, with SteerMusic+ obtaining significantly higher MOS-P and MOS-T than the baselines as measured by an ANOVA test (MOS-P with p-value $1.38\times 10^{-13}$ and MOS-T with p-value $1 \times 10^{-3}$).
In long-form music editing (Tab.~\ref{tab:musicdelta}), SteerMusic+ still outperforms the baseline. We exclude Textual Inv. from comparison as it fails to produce meaningful results on MusicDelta, likely due to the limitation of its base model with complex inputs.
Fig.~\ref{fig:compare1} shows a personalized instrument style transfer example, where SteerMusic+ effectively preserves instruction-irrelevant content from the source music.

\begin{figure}[t]
  \centering
  \includegraphics[width=\linewidth]{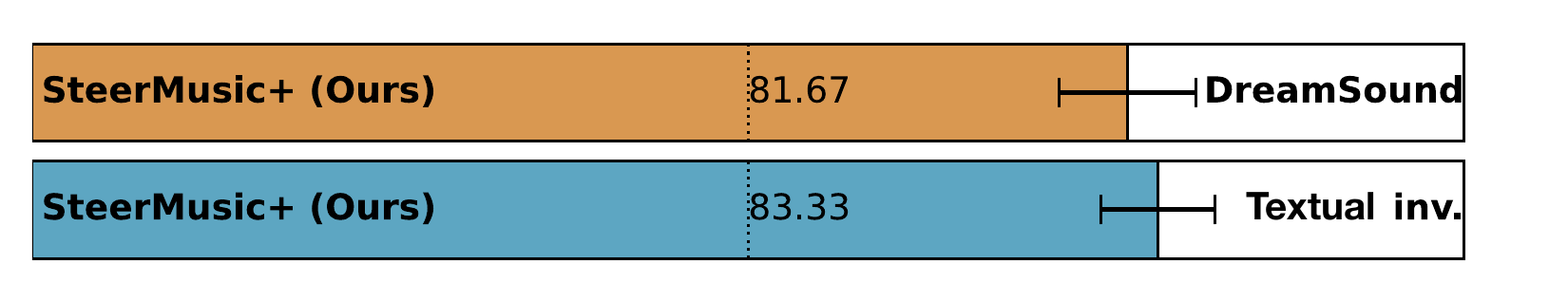}
  \caption{
  User preference for SteerMusic+: percentage of users preferring our method over the baselines, DreamSound and Textual Inv.
  }
  \label{fig:user_study2}
\end{figure}

\textbf{User preference study.} We evaluate SteerMusic+ by a user preference study compared to DreamSound and Textual Inv. Users were asked to select the edited result that best matches the reference style while preserving the main content of the source music. We collected 24 full responses, excluding responses that are partially finished. Participants have a minimum of 1 and an average of 5 years of music training experience. As shown in Fig.~\ref{fig:user_study2}, SteerMusic+ is clearly preferred by participants compared to baselines.

\begin{figure}[t]
  \centering
  \includegraphics[width=\linewidth]{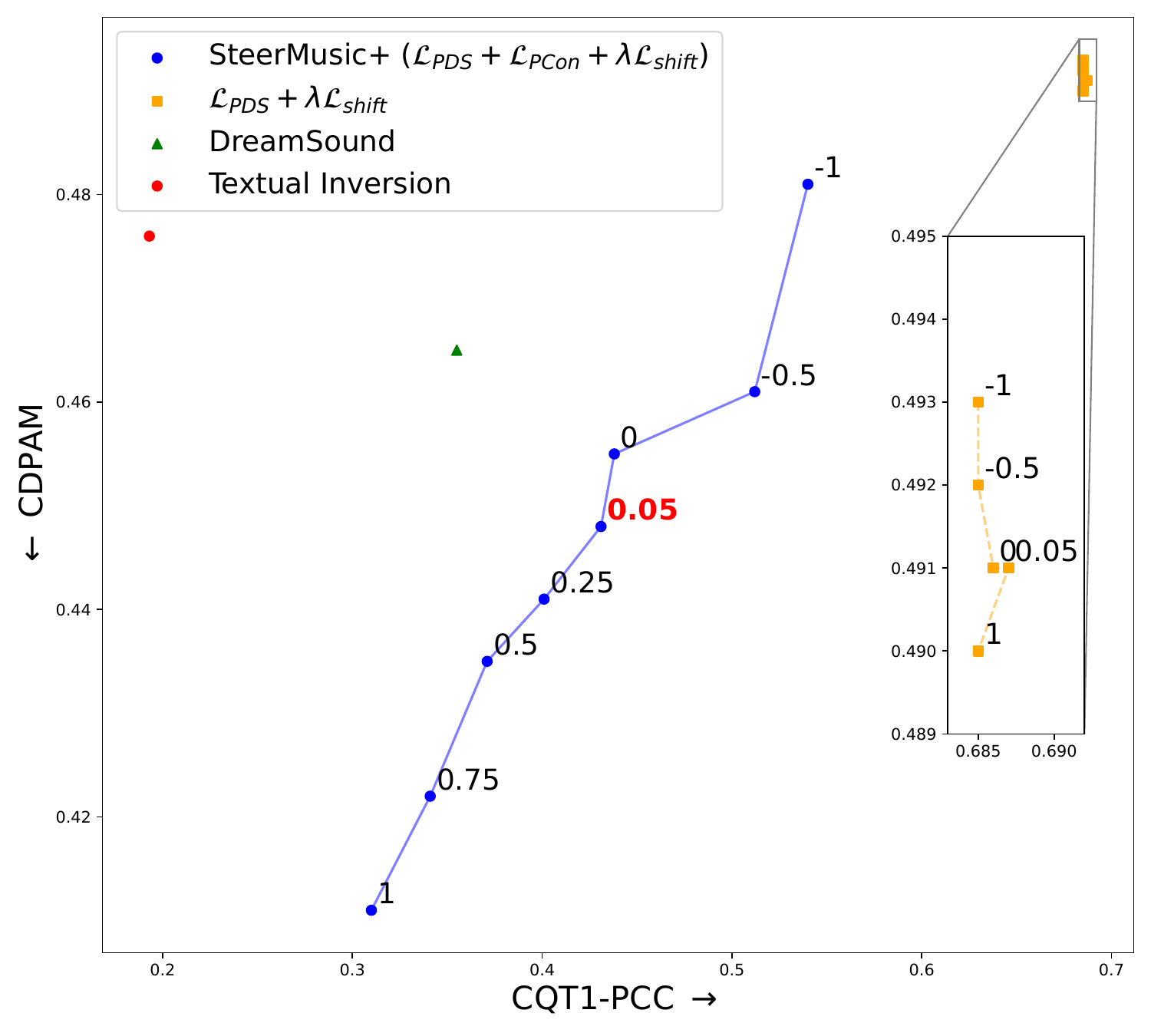}
  \caption{Ablation study on SteerMusic+ with adherence to music consistency vs. edit fidelity on the edited music with $\lambda$ values in Eq.~\ref{eqn:psd_0}. The horizontal axis (CQT1-PCC) indicates source melody preservation; the vertical axis (CDPAM) indicates alignment with the target concept.}
  \label{fig:lambda}
\end{figure}

\textbf{Ablation study.} We conduct an ablation study to understand the effects of different components in SteerMusic+. We used the concept [bouzouki] as it is an uncommon musical instrument that typically requires the use of personalized models. The regularization weight 
$\lambda$, which acts as the Lagrange multiplier for the constraint in Eq.~\ref{eqn:psd_0}, was tested within [-1,1] to balance the constraint enforcement and the optimization stability. Yellow dots indicate results obtained by $\mathcal{L}_{\text{PDS-O}}$ in Eq.~\ref{eqn:psd_0} with varying $\lambda$ values. From the zoomed-in view in Fig.~\ref{fig:lambda}, as the CDPAM value decreases with $\lambda$, the loss of $\mathcal{L}_{\text{shift}}$ within $\mathcal{L}_{\text{PDS-O}}$ helps steer the editing toward the concept. However, the effect of $\mathcal{L}_{\text{shift}}$ is minor, suggesting that the gradient of $\mathcal{L}_{\text{PDS}}$ alone does not sufficiently capture the target concept during editing. As a result, the edited music struggles to align with the reference musical characteristics. 
In contrast, SteerMusic+ (with $\mathcal{L}_{\text{PCon}}$ in Eq.~\ref{eqn:patch_personal}) leads to a significant decrease in the CDPAM value, which indicates better alignment with the target concept. Since $\mathcal{L}_{\text{PCon}}$ was proposed to explicitly enhance the characteristics of the target concept on the editing, we observe that $\mathcal{L}_{\text{shift}}$ becomes more effective in this setting. Specifically, when $\lambda$ is negative, the edit preserves the original content, while positive $\lambda$ pushes the edit toward the target concept.

This highlights a fundamental trade-off between the source music consistency and the adherence to the target style, consistent with the findings of~\citet{manor2024zero}. In personalized music editing, where target characteristics are derived from reference music, shifting the output towards the concept often disrupts the original content. This differs from text-guided editing, which follows abstract textual cues rather than concrete musical references. The key challenge in personalized editing is balancing the retention of the original music content, while integrating the distinctive attributes of the reference music. For practical use, we recommend keeping a smaller $\lambda$ value (e.g., $\lambda=0.05$) to avoid over-editing.

\begin{figure}[t]
  \centering
  \includegraphics[width=\linewidth]{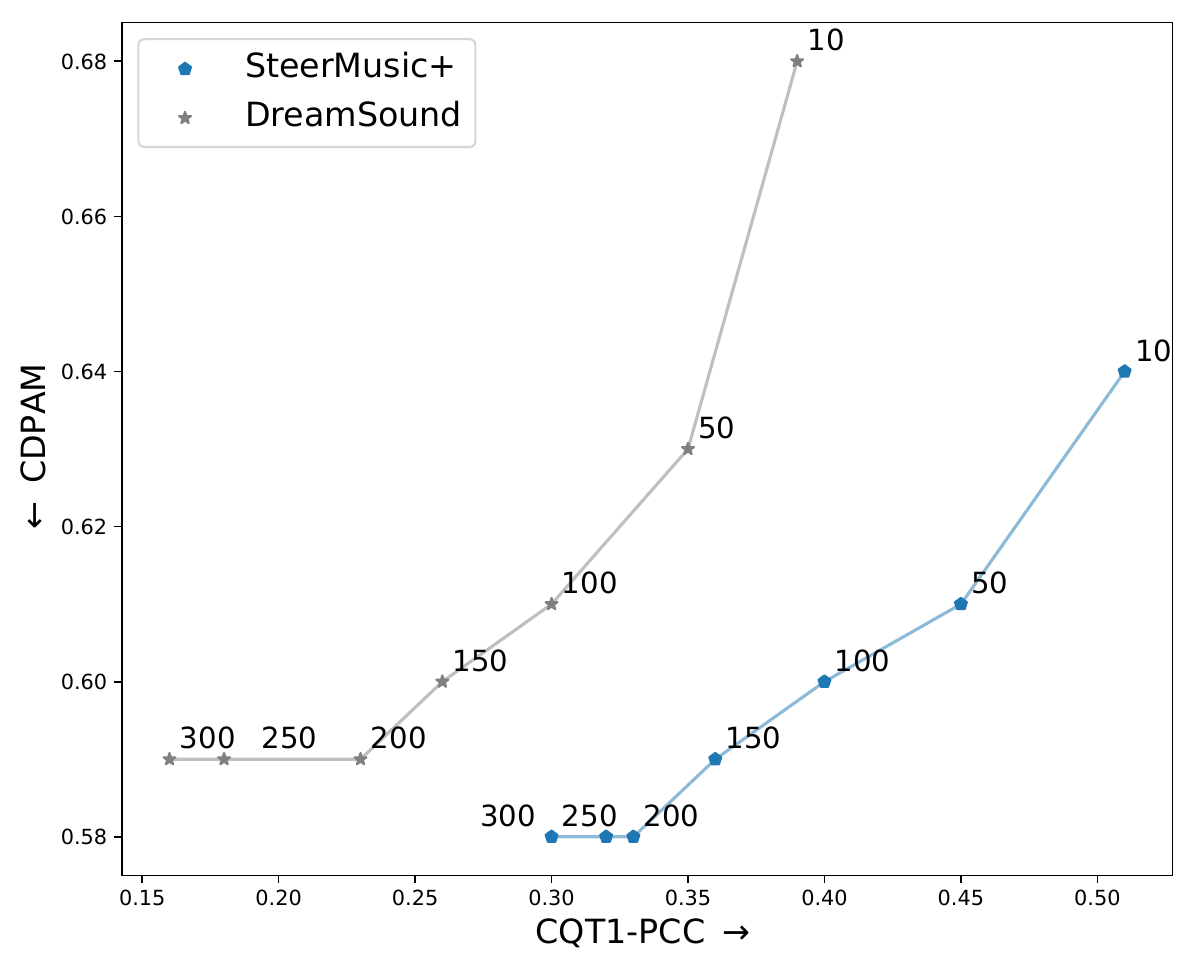}
  \caption{Adherence to music consistency vs. edit fidelity to edited results with different fine-tune steps in PDM. The horizontal axis (CQT1-PCC) indicates source melody preservation; the vertical axis (CDPAM) indicates alignment with the target concept.}
  \label{fig:finetune}
\end{figure}

\textbf{Limitation.} 
As shown in Fig.~\ref{fig:finetune}, the number of fine-tuning steps in the PDM significantly affects the performance of DreamSound and SteerMusic+. A small number of steps (e.g., 50) leads to poor concept learning and low edit fidelity, while a large number of steps (e.g., 200) causes overfitting and source music structural loss. These findings highlight the importance of balancing PDM fine-tuning steps to achieve both high edit fidelity and content preservation in personalized music editing settings. The editing fidelity of SteerMusic+ is fundamentally limited by the ability of the PDMs to capture the concept attributes. Consequently, failure cases can often be attributed to inherent shortcomings in the PDM's representational capacity. We refer readers to \citet{plitsis2024investigating}, which investigated the capacity of PDMs to capture musical attributes and offers insights into improving this capacity.

\section{Conclusion}
We present two music editing methods, \textit{SteerMusic} and \textit{SteerMusic+}, from coarse-grained to fine-grained music editing. To the best of our knowledge, this is the first work to fully leverage DDS and score distillation in the music editing framework. Our methods address the limitations of prior approaches by enhancing musical consistency while producing high-fidelity edits aligned with target prompts. SteerMusic+ further introduces a personalized editing pipeline that extracts user-defined style concepts from reference music for fine-grained control. We validate our methods through comprehensive experiments that show clear improvements over existing baselines in both consistency and fidelity. For practical application, the proposed methods could be extended by using a better-trained TTA diffusion backbone with a higher sampling rate to achieve high-fidelity editing results. In future studies, personalized music editing methods could focus on improving user-controllable editing outcomes for more nuanced and expressive edits.

\bibliography{aaai2026}
\end{document}


\makeatletter
\renewcommand\section{\@startsection {section}{1}{\z@}%
                                   {-2.5ex \@plus -1ex \@minus -.2ex}%
                                   {1.0ex \@plus.2ex}%
                                   {\normalfont\Large\bfseries}}
\renewcommand\subsection{\@startsection{subsection}{2}{\z@}%
                                   {-2.25ex\@plus -1ex \@minus -.2ex}%
                                   {0.75ex \@plus .2ex}%
                                   {\normalfont\large\bfseries}}
\makeatother
\maketitle

\tableofcontents


\section{More Background}\label{supp:more-background}

\subsection{DPMs on Audio Generation}
Given an input audio data $x_0$ and the corresponding text prompt $y$, the forward diffusion process is defined as a Markov chain that gradually adds noise to the input audio data $x_0$ over $T$ steps, and sampled from marginal distribution $q(x_t|x_0):= \mathcal{N}(x_t;\sqrt{\alpha_t}x_0,(1-\alpha_t)I)$ as $x_t = \sqrt{\alpha_t}x_0 + \sqrt{1-\alpha_t}\epsilon_t$, where $\alpha_t$ represents the variance of the forward process at the time step $t$, $\epsilon_t \sim \mathcal{N}(0,I)$ at the time step $t$ and $x_t$ denotes the noisy latent representation of the input $x_0$ at time step $t$. DPMs~\cite{ho2020denoising} learn the backward diffusion process, which denoises from the prior distribution $\mathcal{N}(0,I)$ to the data distribution $x_0$. By predicting the noise $\epsilon_\phi(x_t, y,t)$, the model parameter $\phi$ can be optimized via the following training objective function:
\begin{equation}
\mathcal{L}_{\text{DPM}}(x_0,y; \phi) = \mathbb{E}_{x_t\sim q(x_t|x_0),t\sim \mathcal{U}(1,T)} \| \epsilon_\phi(x_t, y, t) - \epsilon_t \|_2^2   \label{eqn:dpm_loss}
\end{equation}
where $\epsilon_t \sim \mathcal{N}(0,I)$ is the actual noise added to $x_0$, $\mathcal{U}(1,T)$ is a uniform distribution.
In audio generation, the diffusion process can be either performed in data space or in a latent space obtained by VAEs~\cite{kingma2013auto,rombach2022high}. In this study, we perform music editing using AudioLDM2~\cite{liu2024audioldm}, a latent diffusion model (LDM) pretrained on Mel-spectrograms. 
During inference, text prompts are given to AudioLDM2 to generate latent samples. The generated samples are decoded by the pretrained VAE decoder back to the Mel-spectrogram. HiFi-GAN~\cite{kong2020hifi}, a commonly used vocoder, converts the Mel-spectrogram into waveforms.

\subsection{Score Distillation Sampling}
In Score Distillation Sampling (SDS), a pretrained, frozen diffusion model is employed to estimate the score---i.e., the gradient of the log-density---of the conditional distribution \( p(x \mid y) \). The key idea is to optimize a generator function
\[
x = g(\theta)
\]
with respect to \( \theta \) so that the generated data (e.g., an image or an audio) \( x \) attains high likelihood under the diffusion model’s learned density. To this end, we define a differentiable loss \( \mathcal{L}_{\text{SDS}} \) whose minimization produces samples resembling those from the diffusion model. 
\[
\mathcal{L}_{\text{Diff}}(\phi,x=g(\theta)) = w(t)\| \epsilon_\phi (x_t,y,t) - \epsilon\|_2^2\]
In effect, we solve
\[
\theta^* = \arg\min_\theta L_{\text{Diff}}(\phi, x = g(\theta)),
\]
where \( \mathcal{L}_{\text{Diff}}(\phi, x) \) is the original diffusion training loss used to learn \( p(x \mid y) \), and \( \phi \) denotes the parameters of the frozen diffusion model.
 
More precisely, the gradient of the diffusion loss with respect to \( \theta \) is given by
\begin{multline}
\nabla_\theta \mathcal{L}_{\text{Diff}}(\phi, x = g(\theta))
\\ = \mathbb{E}_{\epsilon,t} \left[ w(t) \left(\epsilon_\phi(x_t, y, t) - \epsilon \right) \cdot 
\underbrace{ \frac{\partial\epsilon_\phi(x_t, y, t)}{\partial x} }_{\text{Jacobian}} \cdot 
\frac{\partial x}{\partial \theta} \right].
\end{multline}
Since computing the U-Net Jacobian \( \frac{\partial\epsilon_\phi}{\partial x_t} \) is computationally expensive and poorly conditioned at low noise levels, we omit this term~\cite{poole2022dreamfusion}. The simplified gradient becomes
\[
\nabla_\theta \mathcal{L}_{\text{SDS}}(\phi, x = g(\theta)) \approx \mathbb{E}_{\epsilon,t} \left[ w(t) \left(\epsilon_\phi(x_t, y, t) - \epsilon \right) \cdot 
\frac{\partial x}{\partial \theta} \right]. ~\label{eqn:sds_supp}
\]

Intuitively, this update nudges \( x \) in a direction that increases its (conditional) likelihood according to the diffusion model’s learned score function.

\subsection{Delta Denoising Score}

In image domain, using SDS to perform image editing directly suffers blurry issues~\cite{hertz2023delta}, where the gradient of vanilla SDS can be decomposited into two components:
\begin{equation}
    \nabla_\theta \mathcal{L}_{SDS}(x,y,\epsilon,t) := \delta_{\text{text}} +\delta_{\text{bias}}
\end{equation}
where component $\delta_{\text{text}}$ is a desired direction that directs the optimization to match the condition $y$ (i.e., $y$ is a target prompt in the editing setting), and $\delta_{\text{bias}}$ is undesired component which causes unintended editing on the results such as blurry and smooth.

In the image editing task, given matched and unmatched image-prompt data pairs $\{x^{\text{src}},y^{\text{src}}\}$ and $\{x,y^{\text{tgt}}\}$, respectively. The delta denoising loss can be formulated as 
\begin{multline}
    \mathcal{L}_{\text{DD}}(\phi,x,x^{\text{src}},y^{\text{src}},y^{\text{tgt}}) \\ = \mathbb{E}_{\epsilon,t}[ w(t)\| \epsilon_\phi (x_t,y^{\text{tgt}},t) - \epsilon_\phi (x_t^{\text{src}},y^{\text{src}},t)\|_2^2 ]
\end{multline}
where $x_t$ and $x_t^{\text{src}}$ shares the same sampled noise $\epsilon$. 

Same as in SDS, by omitting the Jacobian over the diffusion model, the gradient over the geneator parameter $\theta$ is given by
\begin{equation}
    \nabla_\theta \mathcal{L}_{\text{DDS}} =\mathbb{E}_{\epsilon,t}[ w(t)(\epsilon_\phi (x_t,y^{\text{tgt}},t) - \epsilon_\phi (x_t^{\text{src}},y^{\text{src}},t))\frac{\partial x}{\partial \theta}] \label{eqn:dds_supp}
\end{equation}
By adding and subtracting $\epsilon$ in Eq.~\ref{eqn:dds_supp}, the DDS can be represented as a difference between two SDS scores:
\begin{equation}
    \nabla_\theta \mathcal{L}_{\text{DDS}} = \nabla_\theta \mathcal{L}_{\text{SDS}}(x,y^{\text{tgt}}) - \nabla_\theta \mathcal{L}_{\text{SDS}}(x^{\text{src}},y^{\text{src}}) ~\label{eqn:dds_sds}
\end{equation}

Thus, \cite{hertz2023delta} claimed the non-zero gradient of the second term in Eq.~\ref{eqn:dds_sds} can be attribured to the noisy direction 
\begin{equation}
    \nabla_\theta \mathcal{L}_{\text{SDS}}(x^{\text{src}},y^{\text{src}}) \approx \delta_{\text{bias}}
\end{equation}
By subtracting the bias term, DDS can be considered a distilled direction that concentrates on editing the relevant portion of the inputs (i.e., image) to match to the target prompt $y^{\text{tgt}}$.

\subsection{Denoising Diffusion Implicit Model}

Given a diffusion probabilistic model parameterized by $\phi$ and a diffusion process defined as $q(x_t|x_0) := \mathcal{N}(x_t;\sqrt{\alpha_t}x_0, (1-\alpha_t)I)$, where the $\alpha_t$ represents the variance of the forward diffusion process at time step $t$, $x_t$ represents the noised latent representation of the data $x_0$. The DDIM~\cite{song2020denoising} defines a update rule in the reverse diffusion process, which the formulation is given by
\begin{multline}
    x_{t-1} =  \underbrace{\sqrt{\alpha_{t-1}}(\frac{x_t-\sqrt{1-\alpha_t}\epsilon^{(t)}_\phi(x_t)}{\sqrt{\alpha_t}})}_{\text{`predicted $x_0$'}} \\+\underbrace{\sqrt{1-\alpha_{t-1}-\sigma^2_t}\epsilon_\phi^{(t)}(x_t)}_{\text{`direction pointing to $x_t$'}}+\underbrace{\sigma_t\epsilon_t}_{\text{`random noise'}}
\end{multline}
where $\sigma_t$ is a free variable that controls the stochasticity in the reverse process.

\textbf{DDIM Inversion.}
By setting $\sigma_t$ to $0$, we can obtain a deterministic update rule which can be reversed to a deterministic mapping between $x_0$ and its latent representation $x_T$. The inverse mapping is refered as DDIM inversion, which is formulated as 
\begin{equation}
    \frac{x_{t+1}}{\sqrt{\alpha_{t+1}}}-\frac{x_t}{\sqrt{\alpha_t}} = (\sqrt{\frac{1-\alpha_{t+1}}{\alpha_{t+1}}}-\sqrt{\frac{1-\alpha_t}{\alpha_t}})\epsilon_\phi^{(t)}(x_t)
\end{equation}

\subsection{Classifier-free Guidance}
Given a diffusion model jointly trained on conditional and unconditional embeddings. In the sampling phase, samples can be generated using classifier-free guidance (CFG)~\cite{ho2021classifier}. The prediction with the conditional and unconditional estimates are defined as following equation
\begin{equation}
    \epsilon_\phi^{\omega}(x_t,h_t):= \omega \epsilon_\phi(x_t,y,t)+(1-\omega)\epsilon_\phi(x_t,\varnothing,t)
\end{equation}
where $\omega$ is the guidance scale that controls the trade-off between mode converage and sample fidelity, and $\varnothing$ is a null token used for unconditional prediction.

\section{User Study}

This section provides detail of our user study design. We design anonymous surveys and only collect responses from participants over 18 years old with at least one year of musical training background. Listening studies have been approved by the privacy committee to ensure ethical compliance with local regulations.

\subsection{User Preference Study for Model Comparison}~\label{sec:preference_study}

\begin{figure*}[h]
  \centering
  \includegraphics[width=0.7\textwidth]{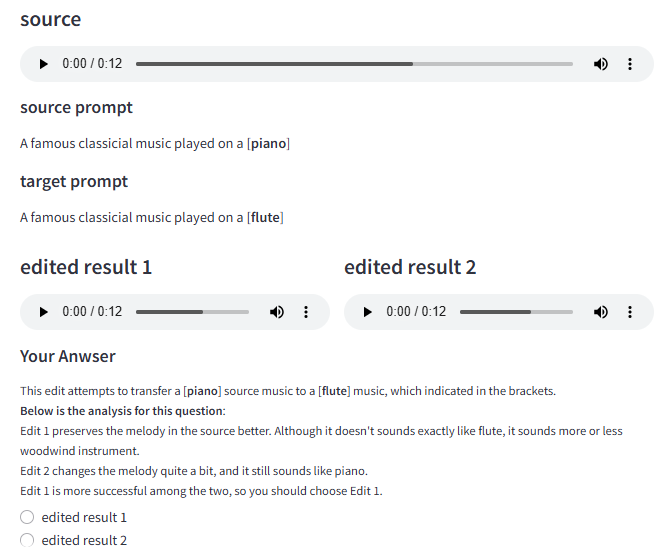}
  \caption{The tutorial for a sample question before the text-guided music editing listening test.}
  \label{fig:tutorial1}
\end{figure*}
\begin{figure*}[h]
  \centering
  \includegraphics[width=0.7\textwidth]{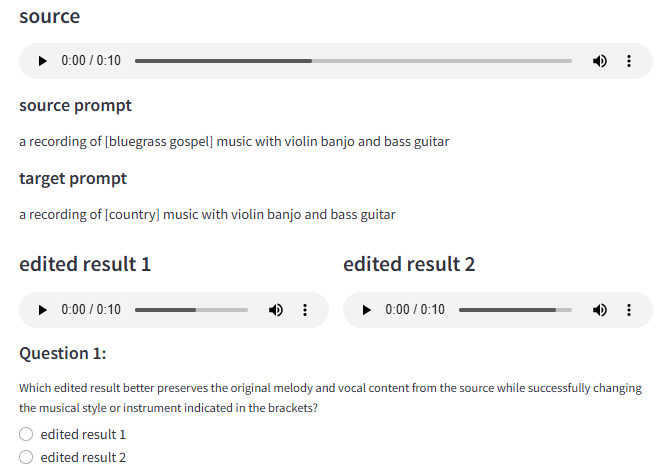}
  \caption{A sample question for the text-guided music editing listening test.}
  \label{fig:sample1}
\end{figure*}

\begin{figure*}[h]
  \centering
  \includegraphics[width=0.7\textwidth]{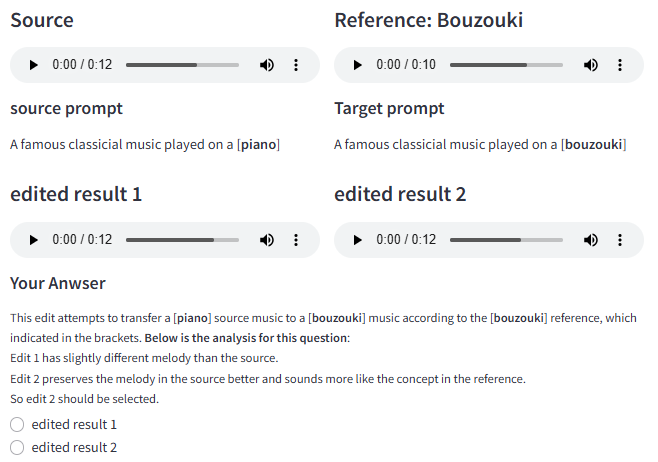}
  \caption{The tutorial for a sample question before the personalized music editing listening test.}
  \label{fig:tutorial2}
\end{figure*}

\begin{figure*}[h]
  \centering
  \includegraphics[width=0.7\textwidth]{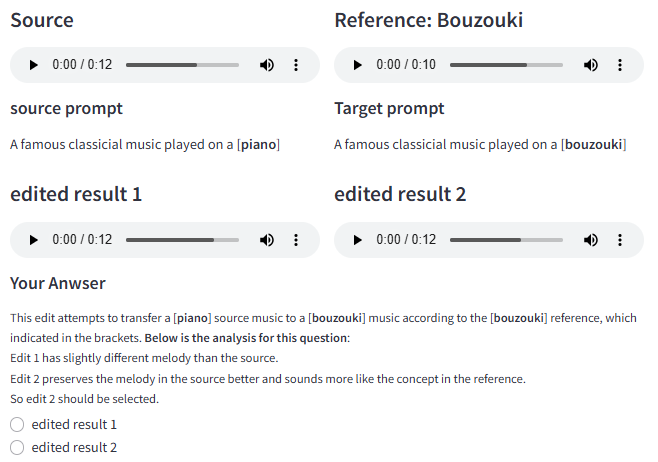}
  \caption{A sample question for the personalized music editing listening test.}
  \label{fig:sample2}
\end{figure*}

We follow the design of~\cite{manor2024zero} for this user preference study. This user preference study contains two parts, the first part is for evaluating text-guided music editing methods and the second part is for evaluating the personalized music editing methods.
We randomly select 10 source musics with corresponding source and target prompts from ZoME-Bench~\cite{liu2024medic} dataset in this user study, each music has 10 seconds duration. For each question, we provide two edited music, one is obtained by our method and the other one is obtained by the compared method, users are asked to select the best matched edited music according to the question. We distribute this user study questionnaire to some open-public groups who are interested in music and have at least one year music training. The order of questions and edited samples are also randomly shuffled in our questionnaire.

For the first part, we include 20 questions with 10 source musics and compare with two methods (i.e., MusicMagus~\cite{zhang2024musicmagus} and ZETA~\cite{manor2024zero}). For each question, we provide a source music, a edit instruction and two edited results. We ask users to select the best-matched result from the two provided results according to the question. Figure~\ref{fig:tutorial1} and Figure~\ref{fig:sample1} demonstrate the tutorial to guide users to select the best matched choice before the main listening test of text-guided music editing, and an exact sample question in this user study.

For the second part, we include 20 questions with 10 source musics and compare with two methods (i.e., DreamSound~\cite{plitsis2024investigating} and Textual inversion~\cite{plitsis2024investigating}). For each question, we provide a source music, a edit instruction, a reference music for the target style,  and two edited results. We ask users to select the best-matched result from the two results provided according to the question. Figure~\ref{fig:tutorial2} and Figure~\ref{fig:sample2} demonstrate the tutorial to guide users to select the best matched choice before the main listening test of personalized music editing, and an exact sample question in this user study.

This user study is anonymous, before the user study, participants were asked to provide their age and number of years for music training.


\subsection{Mean Opinion Score Study}
In order to test the objective metric sensitivity, we conduct additional mean opinion score (MOS) study to further verify our method compared to the baselines for source music correspondence and target style consistency subjectively. Similar as the user preference study in Section~\ref{sec:preference_study}, the MOS study contains two parts: The first part is to verify SteerMusic with 4 baselines (DDIM~\cite{song2020denoising}, SDEdit~\cite{meng2021sdedit}, ZETA~\cite{manor2024zero}, and MusicMagus~\cite{zhang2024musicmagus}), which contains 5 randomly selected source music with edited results. The second part is to verify SteerMusic+ with 2 baselines (Textual inv. and DreamSound~\cite{plitsis2024investigating}), which contains 5 randomly selected source music with edited results. Each music sample has 10-second duration; the MOS study test takes approximately 15 minutes to be completed. The order of questions and edited samples are randomly shuffled in our questionnaire. We distribute this user study questionnaire to some open-public groups who are interested in music and have at least one year of music training. 

\begin{figure*}[t]
  \centering
  \includegraphics[width=0.7\textwidth]{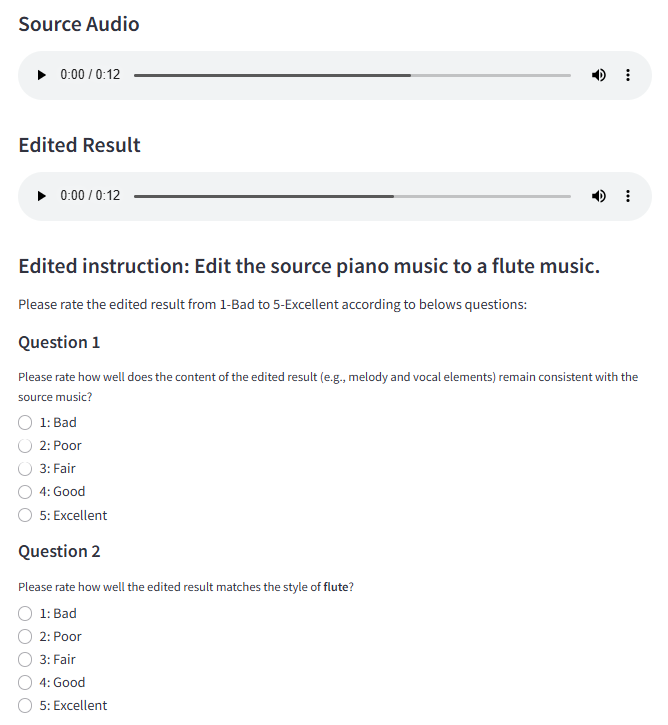}
  \caption{A sample question for MOS study test (Part1) for SteerMusic.}
  \label{fig:MOS1}
\end{figure*}

\begin{figure*}[t]
  \centering
  \includegraphics[width=0.7\textwidth]{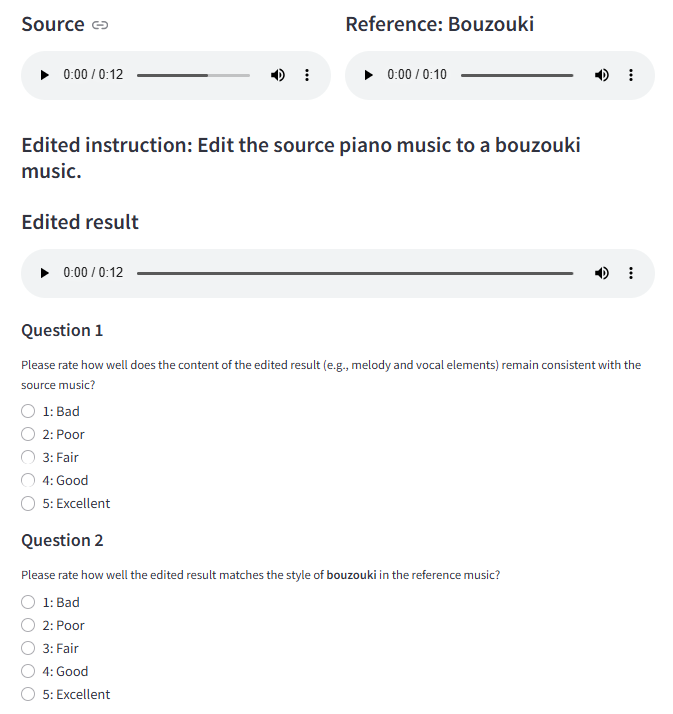}
  \caption{A sample question for MOS study test (Part2)  for SteerMusic+.}
  \label{fig:MOS2}
\end{figure*}

Each of the edited results is followed by two questions:
\begin{enumerate}
    \item Please rate how well the content (e.g., melody and vocal elements) remains consistent with the source music.
    \item Please rate how well the edited result matches the target style.
\end{enumerate}
Participants were asked to give their rate from 1- Bad to 5-Excellent. Example questions for part 1 and part 2 can be found in Figure~\ref{fig:MOS1} and Figure~\ref{fig:MOS2}. 
We collected 23 complete responses for Part 1 and 20 full responses for Part 2 from participants with at least 1 year and on average 3 years of music training experience.

\section{Experimental Details}

\subsection{Experimental Setup}

To evaluate our methods, we use a pretrained AudioLDM2~\cite{liu2024audioldm} as the foundational model. In the text-guided music editing experiment, we set the optimization iteration as 400 with a 30 guidance scale~\cite{ho2021classifier} in the diffusion model in SteerMusic. In personalized music editing experiment, we follow the setting of the official repository of DreamSound\footnote{\label{foot:dreamsound}\url{https://github.com/zelaki/DreamSound}} and fine-tune the DreamSound model 100 steps with $1e^{-5}$ learning rate. We set 400 optimization steps with 15 guidance scale and 0.05 $\lambda$ value in PDS-O equation of SteerMusic+. All experiments were performed on a single NVIDIA H100 GPU.

\subsection{Evaluation Metrics}

In our experiment for zero-shot text-guided music editingt task, we follow~\cite{manor2024zero,copet2023simple,gui2024adapting} , and use the "music\_audioset\_epch\_15\_esc\_90.14.pt" checkpoint of LAION-AI~\cite{chen2022hts,wu2023large} to calculate the CLAP score between target prompts and edited music. Since ZoME-banch~\cite{liu2024medic} dataset contains music clips with 10-second duration, and since this checkpoint was trained for 10-second long segments. We do not apply windows when calculating the CLAP score.

We use CQT2010 function in nnAudio library~\footnote{\url{https://github.com/KinWaiCheuk/nnAudio}} to calculate CQT features, where we set n\_bins = 128 an bins\_per\_octave=24 under 16000 Hz sampling rate. For the CQT-1 PCC metric, we follow~\cite{hou2024editing} and extract the top 1 CQT bins where contains the most of melody information. The detail CQT-1 PCC metric can be formulated as
\begin{equation}
    \text{CQT-1 PCC} =\frac{\sum_{i}^T({c}^{src}_i - \Bar{c}^{src})(c^{tgt}_i-\Bar{c}^{tgt})}{\sqrt{\sum_i^T ({c}^{src}_i - \Bar{c}^{src})^2 \sum_i^T (c^{tgt}_i-\Bar{c}^{tgt})^2}}
\end{equation}
where $c_i$ is the $i$th index of CQT-1 value.

To calculate FAD scores, we obtained FAD$_{\text{Vggish}}$ scores using audioldm-eval in \url{https://github.com/haoheliu/audioldm_eval} based on NumPy version 1.24.4. We found this toolkit with the Numpy version 1.23.5 will return different levels of results. To calculate FAD$_{\text{CLAP}}$, we use the fadtk toolkit in \url{https://github.com/microsoft/fadtk}.

\section{Additional Experimental Results}

\subsection{Significant Test on MOS Study Results}

We conducted significance tests on the MOS scores collected from our user study. In the zero-shot text-guided music editing experiment, we obtained a p-value of 7.37e-27 for MOS-P and 3.24e-7 for MOS-T. In the personalized music editing experiment, the p-values were 1.38e-13 for MOS-P and 1e-3 for MOS-T. All p-values are significantly smaller than 0.05, indicating that the improvements of our proposed methods in both MOS-P and MOS-T are statistically significant. 

\subsection{Quantify DDIM inversion error on Music samples}

We performed 20 steps DDIM inversion on ZoME-bench change instrument subset and obtained 0.077 FAD$_{\text{CLAP}}$ and 0.477 CQT-1 PCC. These metrics further indicate the reconstructed results are different from the original source music.

\subsection{Detail Experimental Results of SteerMusic+ Cross Music Concepts}
Table~\ref{tab:SteerMusic+_instrument} and Table~\ref{tab:SteerMusic+_style} provide detailed results of the model comparison for different concepts of musical instruments and music genre. According to the tables, SteerMusic+ outerperforms the baseline methods cross different musical concepts, indicating its superiors for a higher edit fidelity on personalized music editing with enhanced instruction-irrelevent source music content consistency. 

In Table~\ref{tab:SteerMusic+_instrument}, we include an extra objective metric that calculates cut-off MFCCs cosine similarity (MFCCs COS) between edited music and reference music. Following~\cite{cifka2021self}, we design this metric as additional objective metric to evaluate perceptual timbre similarity between edited results and reference music, where the metric is given by
\begin{equation}
    \text{MFCCs COS} = \text{cos}(f_{c:13}^{\text{tgt}},f_{c:13}^{\text{ref}})
\end{equation}
where $f$ is a cut-off MFCCs feature of musical signal $x$, $c$ represents the cut-off frequency bins. We set $c = 3$ in our experiment. By excluding the lower frequency bins of the MFCCs, which primarily capture pitch and note-related information, the higher frequency bins can be emphasized to better capture timbre characteristics. The MFCCs COS metric can potentially measures the timbre similarity.

In Table~\ref{tab:SteerMusic+_musicdelta}, we provided detail experiment results across the eight concepts on the long-form music dataset, the MusicDelta.

\subsection{More experiment results and visualization for SteerMusic}

\begin{figure*}[t]
  \centering
  \includegraphics[width=0.98\linewidth]{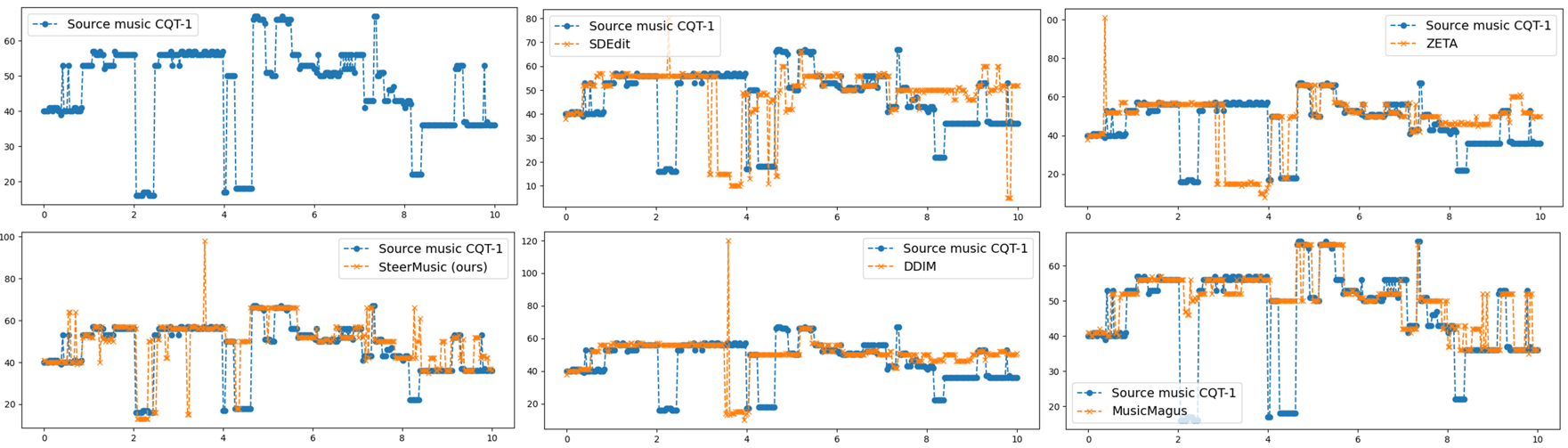}
  \vskip -0.12in
  \caption{A top CQT feature comparison between source and text-guided edited music for different methods. \textnormal{SteerMusic (ours) produces edited results have similar top CQT compare to other methods, indicating SteerMusic successfully preserves source music melody.}}
  \label{fig:CQT_compar_1}
  \vskip -0.1in
\end{figure*}

\begin{figure*}[t]
  \centering
  \includegraphics[width=\linewidth]{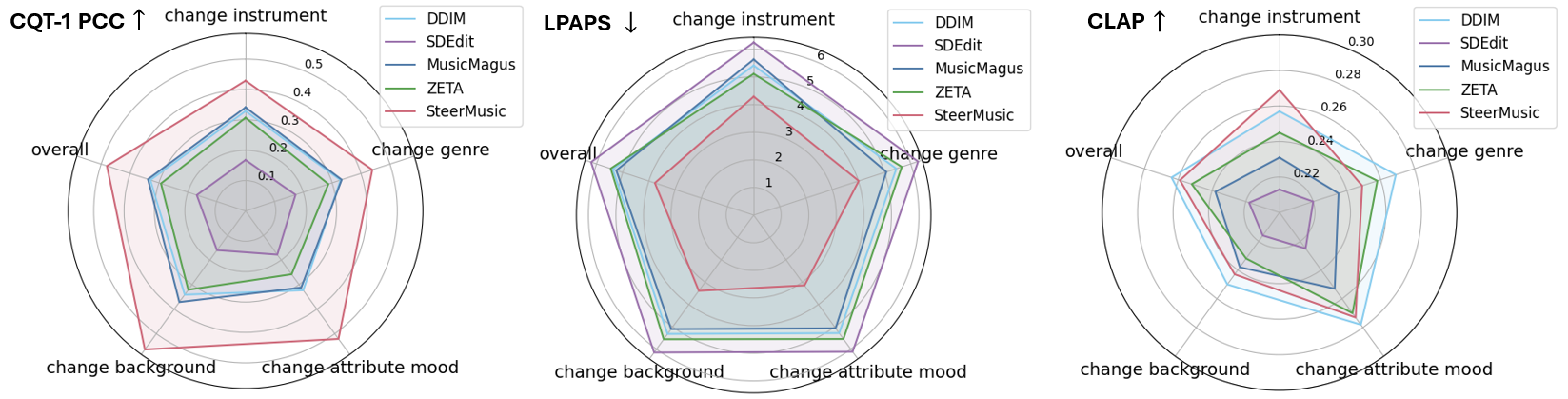}
  \vskip -0.15in
  \caption{Radar chart of model comparison for each text-guided music editing subtask on SteerMusic.}
  \label{fig:spider}
  \vskip -0.15in
\end{figure*}

Fig.~\ref{fig:CQT_compar_1} presents a visualized comparison of top-1 CQT features for an instrument editing sample. SteerMusic generates results with CQT features that closely resemble those of the original music, suggesting that it effectively preserves the original music melody consistency. A more detailed comparison across individual subtasks is in Fig.~\ref{fig:spider}.

\subsection{More visualization for SteerMusic+}

\begin{figure*}[t]
  \centering
  \includegraphics[width=\textwidth]{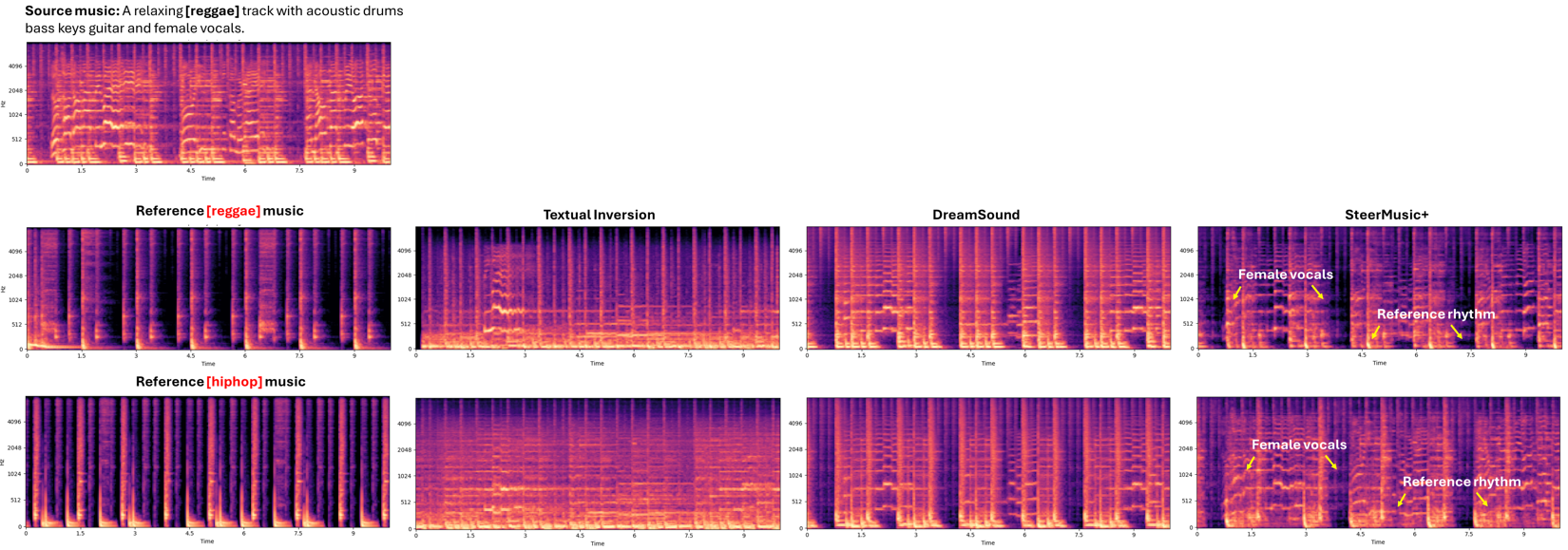}
  \caption{More visualization comparison between SteerMusic+ and baseline methods on personalized genre transfer. \textnormal{SteerMusic+ successfully preserve the vocal content in the source music while perform precise personalized genre transfer.}}
  \label{fig:genre_transfer_1}
\end{figure*}

Figure~\ref{fig:genre_transfer_1} presents an additional visual comparison between SteerMusic+ and other baseline methods (DreamSound and Textual inversion) across various musical style concepts on the same source music, further highlighting the superiority of SteerMusic+ in preserving music content while achieving high edit fidelity aligned with the target concept.

\section{Classifier-Free Guidance Strength v.s. Algorithm Efficiency}

In this section, we conduct an additional ablation study on classifier-free guidances (CFG) strength, which serves as a important hyper-parameter in SteerMusic and SteerMusic+.

\begin{figure*}
  \centering
  \includegraphics[width=0.5\textwidth]{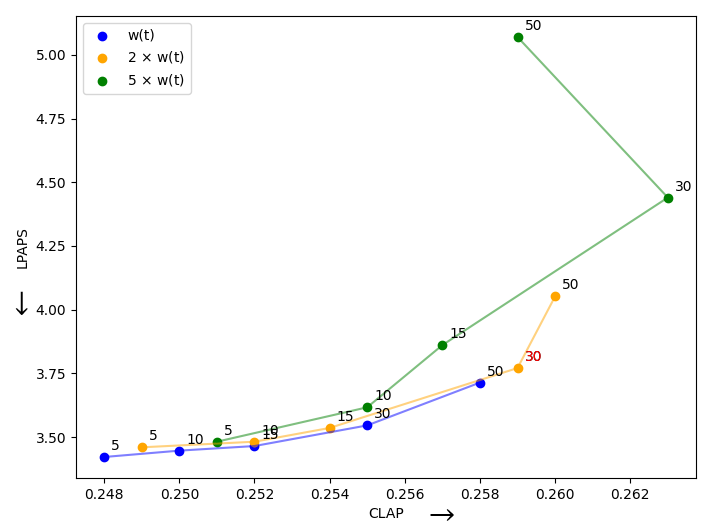}
  \caption{Ablation study of SteerMusic analyzing the trade-off between style correspondence (CLAP) and source music content consistency (LPAPS) under varying classifier-free guidance (CFG) values under 400 optimization steps.\textnormal{ Results are shown for three levels of weight scaling on the weighting function $w(t)$: $1\times$, $2\times$, and $5\times$. Increasing CFG improves alignment with the target prompt (higher CLAP) but often at the cost of higher LPAPS, indicating reduced structural fidelity to the source. In our experiment, we use CFG=30 with 2 times $w(t)$.} }
  \label{fig:cfg}
\end{figure*}

\subsection{CFG Strength for SteerMusic}

Following the observations in~\citet{hertz2023delta}, where higher classifier-free guidance (CFG) values lead to faster optimization convergence, we conducted an ablation study on SteerMusic by varying CFG values and DDS gradient scales, as illustrated in Figure~\ref{fig:cfg}. All experiments were conducted with 400 optimization steps.

We observe that lower CFG values (e.g., 5) result in lower CLAP scores, particularly when using the same variance scale $ w(t)$, indicating that the edited outputs remain more faithful to the source music. This results in higher consistency but weaker alignment with the target prompt. As the CFG value increases, the model places more emphasis on the target prompt, improving CLAP scores but also increasing LPAPS, which reflects a degradation in structural consistency with the source.

This trade-off becomes more evident under larger DDS gradient scales (e.g.,  5 $\times w(t)$), where the optimization process aggressively deviates from the source content. Although alignment improves, the LPAPS rises sharply, signaling loss of source characteristics. In contrast, moderate CFG values (e.g., 15–30) under lower DDS scales offer a more favorable balance between target-style adaptation and source preservation. However, beyond a certain threshold (e.g., CFG = 50), especially at high gradient scales, the results exhibit signs of over-editing, including sharp increases in LPAPS and instability in content preservation.

We also find that slightly increasing the DDS gradient scale (e.g., 
2 $\times w(t)$) can improve optimization convergence without heavily compromising consistency. These results emphasize the importance of jointly tuning CFG and DDS weights to balance semantic alignment with content preservation in text-guided music editing.

Furthermore, this analysis confirms that CFG and DDS weights significantly affect the optimal number of optimization steps, aligning with the findings of~\citet{hertz2023delta}. We recommend using larger CFG values in conjunction with fewer optimization steps to mitigate over-editing.

\begin{figure*}[t]
  \centering
  \includegraphics[width=\textwidth]{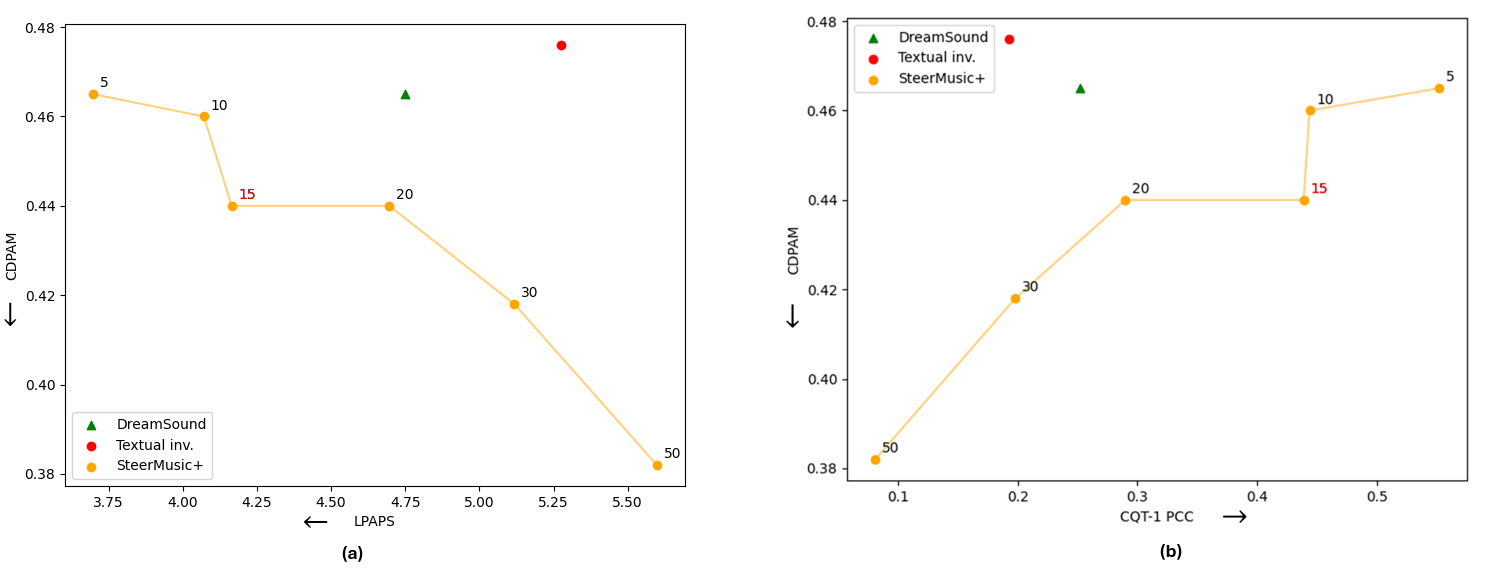}
  \caption{Ablation study of SteerMusic+ analyzing the trade-off between style correspondence and source music melody consistency under varying classifier-free guidance (CFG) values under 400 optimization steps. \textnormal{Increasing CFG values push the edited result closer to the target concept with lower CDPAM; however, it also causes loss source music content (e.g., melody) with higher LPAPS and lower CQT-1 PCC socre.}}
  \label{fig:cfg_steermusicp}
\end{figure*}

\subsection{CFG Strength for SteerMusic+}

In this study, we conduct an ablation study for CFG strength for SteerMusic+ on the personalized music editing task. As shown in Figure~\ref{fig:cfg_steermusicp}, we study how CFG value affects the performance on SteerMusic+. All experiments were run for 400 optimization steps on a personalized diffusion model fine-tuned on [bouzouki] musical concept.

According to Figure~\ref{fig:cfg_steermusicp} (a) and (b), under the same optimization steps, the CFG values controls the closeness of edited results to the target concept as the higher CFG values leading to a lower CDPAM score. However, as we mentioned the experiment section in our main text, it is a trade-off between style consistency and source music content preservation (indicated by CQT-1 PCC values in Figure~\ref{fig:cfg_steermusicp} (a) and LPAPS score in Figure~\ref{fig:cfg_steermusicp} (b)). In our experiment, we set GFG = 15 on SteerMusic+ for the task of personalized music editing. These results highlight the importance of carefully tuning CFG weight scaling to balance semantic alignment and source music content preservation during personalized music editing.

We obtained the same results as the CFG ablation on SteerMusic, where the CFG values affect the optimization strengths given a fixed number of optimization steps. Therefore, we suggest users to set fewer optimization steps when setting a larger CFG values.

\begin{table*}
  \caption{Detail model comparison results across different concepts for personalized music editing task on MusicDelta dataset.}
  \label{tab:SteerMusic+_musicdelta}
  \centering
  \begin{tabular}{c|c|cccccc}
    \toprule
    Method &Concept& FAD$_{\text{CLAP}}$ $\downarrow$ &FAD$_{\text{Viggish}}$ $\downarrow$& CQT-1 PCC $\uparrow$ & LPAPS $\downarrow$ & CDPAM $\downarrow$ \\
    \midrule
    DreamSound & Bouzouki & 0.885 & 9.722 &  \textbf{0.224} & 5.037 & 0.440\\   
    SteerMusic+ & Bouzouki & \textbf{0.855} & \textbf{7.601} & 0.197 & \textbf{4.770} & \textbf{0.426}\\  
    \hline
    DreamSound & Sitar &  \textbf{0.789} & 11.191 & 0.091 & 5.536 & 0.235\\   
    SteerMusic+ & Sitar & 0.961 & \textbf{7.369} & \textbf{0.189} & \textbf{5.289} & \textbf{0.191}\\  
    \hline
    DreamSound & Ocarina &  0.865 & 9.764 & 0.109 & 5.533 & \textbf{0.721}\\   
    SteerMusic+ & Ocarina & \textbf{0.811} & \textbf{6.881} & \textbf{0.324} & \textbf{4.527} & 0.910\\  
    \hline
    DreamSound & Guitar & 0.781 & 8.412 & 0.159 & 4.861 & 0.721\\   
    SteerMusic+ & Guitar &  \textbf{0.535} & \textbf{4.766} & \textbf{0.221} & \textbf{3.848} & \textbf{0.579}\\  
    \hline
    DreamSound & Morricone &  0.859 & 10.269 &  \textbf{0.389} & \textbf{4.875} & 0.478\\   
    SteerMusic+ & Morricone & \textbf{0.642} & \textbf{3.049} & 0.391 & 5.031 & \textbf{0.458}\\  
    \hline
    DreamSound & Reggae & 0.785 & 10.234 & 0.281 & 5.695 &  0.771\\  
    SteerMusic+ & Reggae & \textbf{0.617} & \textbf{5.404} & \textbf{0.344} & \textbf{4.083} & \textbf{0.723}\\ 
    \hline
    DreamSound & Sarabande & 0.930 & 10.269 & 0.234  & 5.468 & \textbf{0.521} \\  
    SteerMusic+ & Sarabande & \textbf{0.365} & \textbf{4.386} & \textbf{0.250} & \textbf{4.440} & 0.582\\
    \hline
    DreamSound & Hiphop & 0.885 & 12.185 & 0.256 &  5.539 & 0.783 \\
    SteerMusic+ & Hiphop & \textbf{0.531} & \textbf{4.594} & \textbf{0.269} & \textbf{4.603} & \textbf{0.782}\\
    \hline
    DreamSound & Overall & 0.847 & 8.972 & 0.220 & 5.318 & 0.583\\
    SteerMusic+ & Overall & \textbf{0.666} & \textbf{5.506} &  \textbf{0.273} & \textbf{4.574} & \textbf{0.581}\\
    \bottomrule
  \end{tabular}
\end{table*}

 \begin{table*}
  \caption{Model comparison on personalized music instrument transfer on ZoME-bench dataset (SteerMusic+ uses the same personalized model as DreamSound).}
  \label{tab:SteerMusic+_instrument}
  \centering
  \begin{tabular}{c|c|cccccccc}
    \toprule
    Method &Concept& FAD$_{\text{CLAP}}$ $\downarrow$&  FAD$_{\text{Viggish}}$ $\downarrow$& CQT-1 PCC $\uparrow$ & LPAPS $\downarrow$ & MFCCs COS $\uparrow$ & CDPAM $\downarrow$ \\
    \midrule
    Textual inv. & Guitar & 0.565 & 7.660 & 0.148 & 5.341 & \textbf{0.666} & 0.803 \\
    DreamSound & Guitar & 0.683 & 9.432  & 0.247 & 4.949 & 0.647 &  0.739 \\
    SteerMusic+ &  Guitar & \textbf{0.358} & \textbf{4.402} & \textbf{0.425} & \textbf{3.963} & 0.637 & \textbf{0.711}\\
    \hline
    Textual inv. & Ocarina & 0.490 & 15.175 & 0.184 & 5.261 & \textbf{0.094} & 0.988\\
    DreamSound & Ocarina  & 0.714 & 7.252 & 0.347 & 4.976 & -0.097 & 0.922\\
    SteerMusic+ & Ocarina & \textbf{0.341} & \textbf{3.702} &  \textbf{0.493} & \textbf{3.913} & 0.045 & \textbf{0.919}\\
    \hline
    Textual inv. & Bouzouki & 0.450 & 6.127 & 0.193 & 5.274 & 0.576 &  0.476\\
    DreamSound & Bouzouki & 0.577 & 8.639 & 0.355 & 4.750 & 0.761 & 0.464 \\
    SteerMusic+ & Bouzouki & \textbf{0.358} & \textbf{5.172} & \textbf{0.439} & \textbf{4.165} & \textbf{0.773} & \textbf{0.440}\\
    \hline
    Textual inv.& Sitar & 0.526 & 5.893 & 0.206  & 5.218 & 0.297 & 0.376\\
    DreamSound & Sitar & 0.770 &  4.741 & 0.230 & 5.303  & 0.772 & 0.269 \\
    SteerMusic+ & Sitar & \textbf{0.450}&  \textbf{3.927} & \textbf{0.266} & \textbf{4.509} & \textbf{0.830} & \textbf{0.229}\\
    \bottomrule
  \end{tabular} 
\end{table*}

\begin{table*}
  \caption{Model comparison on personalized music genre transfer on ZoME-bench dataset (SteerMusic+ uses the same personalized model as DreamSound)}
  \label{tab:SteerMusic+_style}
  \centering
  \begin{tabular}{c|c|cccccc}
    \toprule
    Method &Concept& FAD$_{\text{CLAP}}$ $\downarrow$&  FAD$_{\text{Viggish}}$ $\downarrow$& CQT-1 PCC $\uparrow$ & LPAPS $\downarrow$ & CDPAM $\downarrow$ \\
    \midrule
    Textual inv.& Morricone & 0.496 & 16.334 & 0.253 & 4.815& 0.609\\
    DreamSound & Morricone &  0.720 & 19.128 & 0.289 & 5.093 & 0.469\\   
    SteerMusic+ & Morricone & \textbf{0.312} & \textbf{3.714} & \textbf{0.459} & \textbf{3.896} & \textbf{0.465}\\  
    \hline
    Textual inv. & Reggae& 0.446 & \textbf{5.377} & 0.199 & 5.062 & 0.804\\ 
    DreamSound & Reggae & 0.657 & 10.831 & 0.312 & 5.309  & \textbf{0.700}\\  
    SteerMusic+& Reggae & \textbf{0.432} & 6.040  & \textbf{0.319} & \textbf{4.416} & 0.705 \\
    \hline

    Textual inv.& Sarabande & 0.466& 5.386 & 0.251 & 5.079& 0.815\\ 
    DreamSound & Sarabande & 0.814  & 15.860 & 0.275 & 5.070 & 0.606 \\    
    SteerMusic+ & Sarabande & \textbf{0.333} & \textbf{3.693} & \textbf{0.398} & \textbf{3.997} & \textbf{0.573} \\
    \hline
    Textual inv.& Hiphop & 2.868 &  15.551 & 0.293 & 4.607 &  0.832\\
    DreamSound & Hiphop & 2.280 & 9.207 & 0.268  & 5.209  &  0.702\\
    SteerMusic+ & Hiphop & \textbf{2.078} & \textbf{9.142} & \textbf{0.389}& \textbf{4.139} &  \textbf{0.701}\\
    \bottomrule
  \end{tabular} 
\end{table*}

\section{More Experiment and Discussion for SteerMusic Adaptation}

In this section, we further explore the adaptation of variant score distillation methods within the SteerMusic framework for zero-shot text-guided music editing task. Specifically, we investigate two approaches: the first involves directly adapting the score distillation sampling (SDS) method~\cite{poole2022dreamfusion}, as formulated in Eq.~\ref{eqn:sds_supp}, for zero-shot text-guided music editing. The second approach leverages an improved variant of the DDS method, originally proposed for text-guided image editing, known as Contrastive Denoising Score (CDS)\cite{nam2024contrastive}.

\subsection{Score Distillation Sampling for Zero-shot Text-guided Music Editing}

In our first attempt, we directly adapt vanilla score distillation sampling (SDS)~\cite{poole2022dreamfusion} method for text-guided music editing, which the gradient over $\theta$ is given by 
\begin{equation}
    \nabla_\theta  \mathcal{L}_{\text{SDS}} (x,y^{\text{tgt}},\epsilon,t) = \mathbb{E}_{\epsilon,t}[w(t)(\epsilon_\phi (x_t,y^{\text{tgt}},t)-\epsilon)\frac{\partial x}{\partial \theta}]
\end{equation}
where $\epsilon\sim \mathcal{N}(0,I), t\sim \mathcal{U}(1,T)$.

\begin{table*}[t]
  \caption{Model comparison between SteerMusic and other score distillation adaptation methods on different music style transfer sub-tasks. \textnormal{SteerMusic$^\diamond$ represents the results with extra $ \mathcal{L}_{\text{PatchNCE}}(x,x^{src})$ defined in Eq.~\ref{eqn:patch} in the SteerMusic.}}

  \label{tab:SteerMusic}
  \centering
  \begin{tabular}{c|c|cccccccc}
    \toprule
    Method & Task  & FAD$_{\text{CLAP}}$ $\downarrow$ &FAD$_{\text{Viggish}}$ $\downarrow$ & CQT-1 PCC$\uparrow$&  CLAP $\uparrow$& LPAPS $\downarrow$\\
    \midrule
    SDS & Change instrument& 2.178  & 6.886 &  0.294 & 0.267 & 4.938 \\
    SteerMusic & Change instrument & \textbf{0.257} &  3.005 & 0.429 & \textbf{0.269} & 4.291 \\
    SteerMusic$^\diamond$  & Change instrument & 0.277  & \textbf{1.215} & \textbf{0.685} & 0.236 & \textbf{3.435}\\
    \hline
    SDS & Change genre & 2.529& 7.649 & 0.233 & \textbf{0.268} & 5.028\\
    SteerMusic & Change genre & 0.278 & 2.902 & 0.439  & 0.249& 4.013 \\
    SteerMusic$^\diamond$  & Change genre & \textbf{0.259} & \textbf{1.229} &  \textbf{0.647} & 0.221 &   \textbf{3.474}\\
    \hline
    SDS & Change mood & 2.801 & 5.133 & 0.284 & \textbf{0.277} & 4.784\\
    SteerMusic  & Change mood & 0.275 & 1.607  & 0.521 & 0.273 &  \textbf{3.145}\\
    SteerMusic$^\diamond$  & Change mood & \textbf{0.273} &\textbf{1.187} &  \textbf{0.644} &  0.272 & 3.396\\
    \hline
    SDS & Change background & 2.152 & 5.966& 0.273 & \textbf{0.268} & 4.877 \\
    SteerMusic & Change background &  0.312 & 1.830 & 0.564 & 0.243 & 3.402\\
    SteerMusic$^\diamond$  & Change background & \textbf{0.310} & \textbf{0.998} & \textbf{0.702} & 0.242 & \textbf{3.388}\\
    \hline
    SDS & Overall & 2.410 & 6.534 & 0.270 & \textbf{0.270} & 4.918 \\
    SteerMusic & Overall & \textbf{0.278} & 2.426 & \underline{0.480} & \underline{0.259} & \underline{3.772} \\
    SteerMusic$^\diamond$  & Overall & \textbf{0.278} & \textbf{1.168} & \textbf{0.669} & 0.241 & \textbf{3.428}\\
    \bottomrule
  \end{tabular}
\end{table*}

\subsection{SteerMusic with Contrastive Loss Regularization}

In our second attempt, we draw inspiration from Contrastive Denoising Score (CDS)~\cite{nam2024contrastive} by incorporating an additional contrastive loss regularization to further enhance source music consistency. The CDS method was originally proposed to solve the limitation of DDS that cannot maintain spatial structure consistency in edited images. We coin the variate SteerMusic method with additional contrastive loss regularization as SteerMusic$^\diamond$.

Inspired by~\cite{nam2024contrastive}, the desired edited results should not only align well with the target prompt, but also incorporating other music structural elements such as melody and harmony of the input source music. Motivated by~\cite{liu2024medic} that uses self-attention queries to refine musical structures during editing. Recent studies in image domain shows that self-attention features of text-to-image diffusion models are embedded with detailed spatial information, which allows to build image semantic correspondence using these features~\cite{yu2025dreamsteerer,alaluf2024cross,tumanyan2023plug,zhang2023tale}. Self-attention features in audio generative diffusion models also indicates an overall audio structures~\cite{liu2024medic}. To this end, we adopt CDS method~\cite{nam2024contrastive} and we include a patchwise contrastive loss between on self-attention features into SteerMusic, which further enhances the source music structures on edited results. 

During DDS gradient computing process, we extract self-attention features as $\hat{h}_l$ and $h_l$, where $h_l$ and $\hat{h}_l$ represents the intermediate features passed through the residual block and self-attention block conditioned on $y^{tgt}$ and $y^{src}$, respectively. Unlike PCon loss in SteerMusic+, we keep the original size of self-attention features which have shape as $\mathbb{R}^{(T_l\times F_l)\times C_l}$, where $T_l,F_l$, and $C_l$ represents the size of temporal, spatial and channel dimension in the $l$-th layer, respectively. The query patch is sampled from the feature map $h_l$. We denote $s \in \{1,2,...,S_l\}$ is the query patch, where $S_l = T_l\times F_l$. For each query, the patch at the corresponding spatial location on the feature map $\hat{h}_l$ is `positive' and the non-corresponding patches within the feature map as `negative'. The positive patch is referred as $\hat{h}^s_l$ and the other patches as $\hat{h}_l^{S_l\backslash s})$. The additional PatchNCE loss function is formally defined as
\begin{equation}
    \mathcal{L}_{\text{PatchNCE}}(x,x^{src}) = \mathbb{E}_h [\sum_l \sum_{s} \ell(h_l^{s},\hat{h}_l^{s},\hat{h}_l^{S_l\backslash s})] \label{eqn:patch}
\end{equation}
\begin{equation}
    \ell(h,h^+,h^-) = -\text{log}(\frac{\text{exp}(h\cdot h^+/\tau)}{\text{exp}(h\cdot h^+/\tau)+ \text{exp}(h\cdot h^-/\tau)})  \label{eqn:contrastive}
\end{equation}
where $\text{exp}(h\cdot h^+/\tau)$ is positive sample that with the same patch location, $\text{exp}(h\cdot h^-/\tau)$ is negative sample with mismatched spatial location in the self-attention features, $\tau$ is a temperature parameter 
as $\tau >0$. Following~\cite{nam2024contrastive}, the gradient of $\mathcal{L}_{\text{PatchNCE}}(x,x^{src})$ loss will propagate to the hidden state of self-attention layers $h$ to regularize $\mathcal{L}_{\text{DDS}}$ to have overall content consistency between $x$ and $x^{src}$.

The function of $\mathcal{L}_{\text{PatchNCE}}(x,x^{src}) $ in SteerMusic is fundamentally different to $\mathcal{L}_{\text{Pcon}}$ loss proposed in SteerMusic+, where in this setting, we calculate contrastive loss between two self-attention features come from the same diffusion model with respect to the spatial location. This additional loss serves the same function as the method proposed by~\cite{nam2024contrastive}, which helps to enhance the source music structure consistency during editing. Since we used a spectrogram-based text-to-audio diffusion model, the source music structture consistency here represents the structure consistency in Mel-spectrogram.

\subsection{Experimental Results}

We make comparison between SteerMusic and the proposed two additional adaptations in above subsections.
Table~\ref{tab:SteerMusic} presents a performance comparison between the original SteerMusic method proposed in the main text and variants, denoted as SteerMusic$^\diamond$ and SDS. SteerMusic$^\diamond$ incorporates an additional contrastive loss introduced by~\cite{nam2024contrastive} to further enhance melody preservation in the source music.

Although SDS achieves the highest CLAP score compared to SteerMusic and SteerMusic$^\diamond$, its significantly lower CQT-1 PCC and LPAPS scores indicate a failure to preserve source music consistency. This result consists to the finding in image editing domain~\cite{hertz2023delta}, which SDS suffers blurry issue and make the edited results difficult to preserve original content. Additionally, SDS yields significantly higher FAD scores, further indicating lower audio quality in the edited results.

In SteerMusic$^\diamond$, the inclusion of the $\mathcal{L}_{\text{PatchNCE}}(x, x^{src})$ loss helps maintain the structural characteristics of the source music in the edited outputs, as evidenced by a higher CQT-1 PCC score and lower LPAPS score. However, this comes at the cost of the lowest CLAP score , suggesting that the edited outputs may be less aligned with the target prompt and produce a failure editing. This implies that SteerMusic$^\diamond$ produces less perceptible edits, leaning the outputs closer to the original music. These results indicate a failed adaptation of the Contrastive Denoising Score (CDS)~\cite{nam2024contrastive}, originally proposed for the image domain, to the music editing task. One possible explanation is that enforcing stronger structural consistency in the Mel-spectrogram constrains frequency-domain edits, leading to reduced editing accuracy. Enforcing structural consistency like $\mathcal{L}_{\text{PatchNCE}}(x, x^{src})$ further push the edited output too close to the source music, suppressing necessary changes in frequency domain, such as timbre and rhythm, that are essential for aligning with the target prompt for style transfer editing. 

Compared to both adaptations, SteerMusic achieves a better balance between source music consistency and edit fidelity, demonstrating its effectiveness in the music editing domain.

\bibliography{aaai2026}